\documentclass[conference]{IEEEtran}
\IEEEoverridecommandlockouts
\usepackage{tikz}
\usetikzlibrary{automata, positioning, arrows, shapes}
\usepackage{algorithm}
\usepackage[noend]{algpseudocode}
\usepackage{amsmath,amssymb,amsfonts}
\usepackage{amsfonts}
\usepackage{graphicx}
\usepackage{comment}
\usepackage{xspace}
\usepackage{caption}
\usepackage{subfig}
\usepackage{float}
\usepackage{url}
\usepackage{tabularx}
\usepackage{enumitem}
\usepackage{mathtools}
\usepackage{cite}
\usepackage{subcaption}
\usepackage{multirow} 
\usepackage{booktabs}
\usepackage{multibib}

\AtBeginDocument{%
  }

\newcommand{\system}{AWaRe-SAC\xspace}

\newcommand\eat[1]{}

\begin{document}

\title{
AWaRe-SAC: Proactive Slice Admission Control under Weather-Induced Capacity Uncertainty
}


\author{Dror Jacoby*, Yanzhi Li*, Shuyue Yu, Nicola Di Cicco, Hagit Messer, Gil Zussman, and Igor Kadota
\IEEEcompsocitemizethanks{\IEEEcompsocthanksitem Dror Jacoby and Hagit Messer are with the School of Electrical Engineering, Tel~Aviv University. Yanzhi Li and Igor Kadota are with the Department of Electrical and Computer Engineering, Northwestern University. Shuyue Yu and Gil Zussman are with the Department of Electrical Engineering, Columbia University. Nicola Di Cicco is with the Department of Electronics, Information and Bioengineering, Politecnico di Milano. This work was supported in part by NSF and Center for Smart Streetscapes (CS3) under NSF Cooperative Agreement EEC-2133516, NSF grant AST-2232455, NSF grant AST-2132700, NSF grant CNS-2433807, NSF grant CNS-2148128 and by funds from federal agency and industry partners as specified in the Resilient \& Intelligent NextG Systems (RINGS) program, and BSF grant T-2025115. 
*These authors contributed equally to this work.}
}

\maketitle 

\begin{abstract}
Millimeter-wave (mmWave) links are increasingly utilized in wireless x-haul transport to meet growing service demands. However, the inherent susceptibility of mmWave links to weather-related attenuation creates uncertainty about future network capacity which can significantly affect Quality of Service (QoS). This creates a critical challenge: how to make admission control decisions for slices with QoS requirements, balancing acceptance rewards against the risk of future QoS-violation penalties due to capacity uncertainty? To address this, we develop a proactive slice admission control framework that tightly integrates: (i) a predictor that leverages historical link measurements to forecast short-term attenuation and quantify uncertainty; and (ii) an admission control algorithm that incorporates both the predictions and uncertainties to maximize rewards and minimize QoS-violation penalties. We compare our framework against baseline, state-of-the-art, and idealized oracle algorithms using real-world mmWave x-haul data and residential traffic traces. Simulations suggest that our framework can achieve revenues that are $250\%$ larger than baseline algorithms and $75\%$ larger than state-of-the-art algorithms.
\end{abstract}


\section{Introduction} \label{sec.intro}
Beyond-5G networks increasingly rely on mmWave frequencies, particularly V Band (57--71 GHz) and E Band (71--76 GHz and 81--86 GHz), for short-range deployments requiring high bandwidth and low latency~\cite{rappaport2013millimeter}. However, these higher frequencies are more susceptible to weather-induced attenuation, with rain causing significant signal degradation, limiting reliability~\cite{nagaraj2023propagation,niu2015survey, ITU}. 
%
%
Satisfying stringent QoS requirements of dynamic traffic demands under time-varying link capacities creates complex challenges for network operators, driving the need for resilient, adaptive strategies to mitigate communication disruptions and optimize network performance.

Most existing slice admission control algorithms (discussed in Sec.~\ref{sec.related}) assume that link capacities are static and known a priori, and, thus, are inadequate for modern mmWave networks. 
In contrast, our work explicitly integrates weather-aware link capacity forecasting and proactive slice admission in mmWave backhaul. Admission decisions account not only for immediate rewards (accrued from slice admission) but also for future penalties arising from potential QoS violations (due to stochastic link capacity degradation).
%
%

\begin{figure}[t]
\centering
\includegraphics[width=\columnwidth]{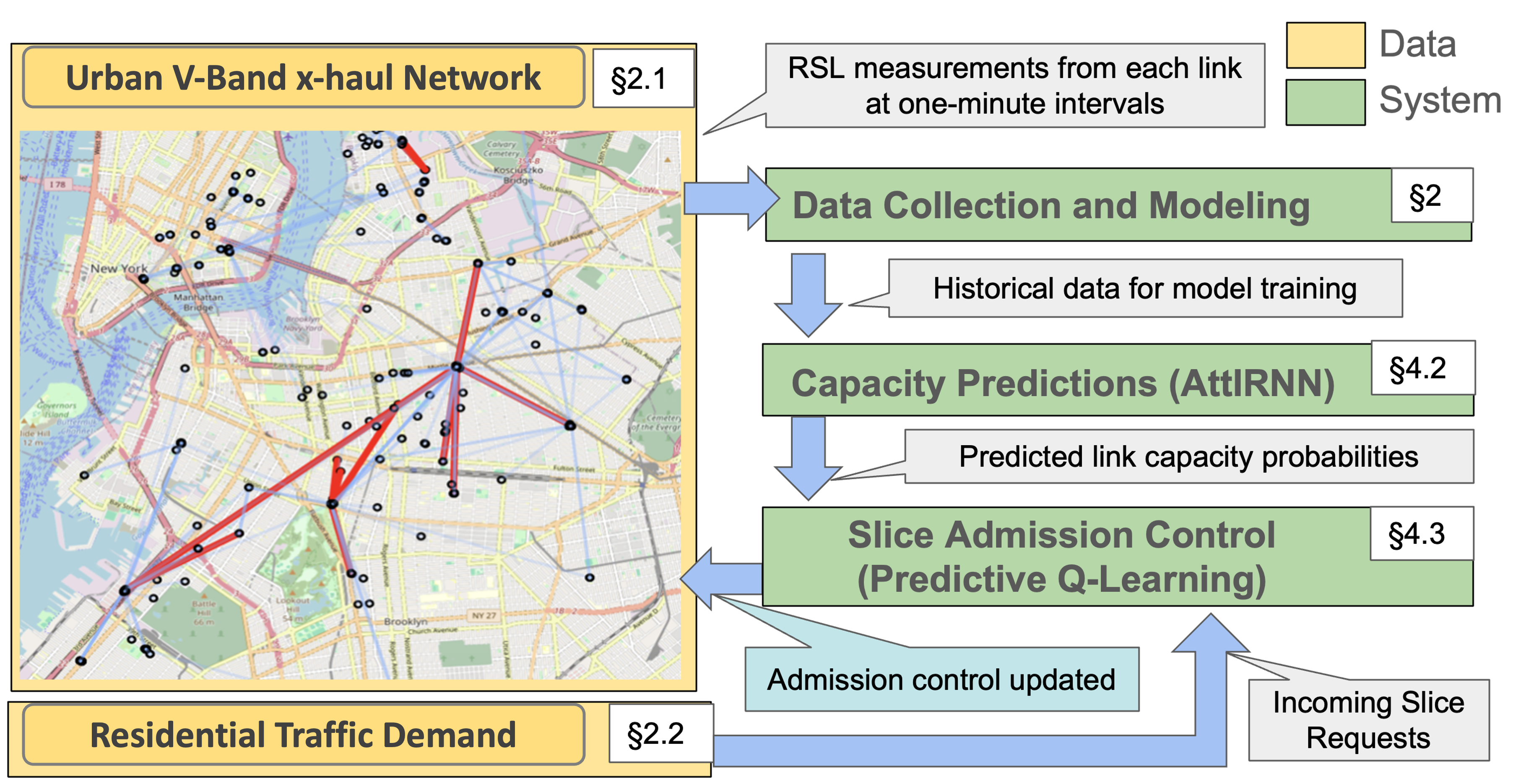}
\caption{Overview of the \system framework workflow.}
\label{fig.design1}
\vspace{-2em}
\end{figure}

We develop an \textbf{A}daptive \textbf{W}eather-\textbf{a}ware \textbf{Re}al-time \textbf{S}lice \textbf{A}dmission \textbf{C}ontrol (\system) framework that tightly integrates: \textbf{(i) Prediction phase} which leverages historical link measurements to forecast short-term link attenuation and quantify estimation uncertainty. \textbf{(ii) Slice admission control phase} which incorporates the sequence of predictions and uncertainties to proactively admit Slice Requests (SRs) aiming to maximize rewards and minimize QoS-violation penalties. 
%
Fig.~\ref{fig.design1} presents the \system framework applied to a segment of the NYC Mesh~\cite{jacoby2025openmeshdata} mmWave network deployed in New York City (NYC). 
Due to space constraints, we use appendices in~\cite{ourArXiv} for details of the mmWave network infrastructure, SR generation mechanism, choice of QoS-violation penalty functions, and more. \textbf{Our contributions include:}
%
\begin{itemize}[leftmargin=0.3cm]
\item \textbf{Slice Admission with Time-varying Resources:} Contrasting with the traditional slice admission problem with static network resources, our network has a unique challenge: QoS-violations may be unavoidable. Our novel problem formulation and \system framework explicitly account for QoS-violation risk and ensuing penalties, including algorithms that address slice underprovisioning. 
\system proactively balances current admission rewards and predicted QoS-violation penalties by tightly coupling forecasting of capacity levels probability distributions with a Q-learning admission control algorithm.

\item  \textbf{Evaluation using Real-World Datasets:} 
We use two datasets: (i) link measurements from the NYC Mesh network, capturing weather-induced attenuation in V Band; and (ii) residential traffic traces from 520 units across 28 buildings in NYC, capturing real user demand patterns. Our predictor leverages the significant short-term temporal correlation of weather-induced attenuation. Slices generated from traces can have longer durations. 
The combination of these datasets provides the first realistic benchmark for slice admission algorithms under weather-induced capacity variations. We validate \system on this benchmark, demonstrating 15–20\% improved prediction accuracy and up to $3.5\times$ higher revenue. 


\end{itemize}

\section{Related Work} \label{sec.related} 
\noindent\textbf{Slice Admission Control.} Prior work on slice admission control explores Reinforcement Learning (RL)-based methods, including tabular Q-learning \cite{bega2017optimising}, deep RL \cite{filali2022dynamic, liu2021onslicing}, and multi-agent methods \cite{sulaiman2022coordinated}, as well as non-RL approaches such as optimization and adaptive methods \cite{wu2022survey, zhao2025adaslicing}. Most prior work on slice admission control assumes static network resources, whereas real-world x-haul networks can be sensitive to weather-induced attenuation, rendering existing solutions inadequate for mmWave networks. Our novel framework not only aims to ``pack'' as many slices as possible in the available resources, but also minimizes the risk of future underprovisioning penalty. Introducing time-varying network resources and underprovisioning penalties, unique to our mmWave x-haul scenario, significantly changes both the constraints and the objective of the admission control problem.

Furthermore, prior work has primarily evaluated admission control algorithms on slices created from simulations or predefined distributions (e.g., Poisson processes) \cite{bega2017optimising, sulaiman2022coordinated}. A main challenge for generating realistic slices is the lack of standardized mechanisms for creating slices from network traces. We address this challenge in Sec.~\ref{sec:slice}. 


\noindent\textbf{Forecasting Network Resources and Traffic Demand.} Traffic demand forecasting has evolved from statistical models to data-driven methods \cite{gutterman2019ran}, with deep learning frameworks explicitly optimizing cost and QoS \cite{bega2019deepcog, collet2022lossleap}. 
A similar trend appears in rain-induced attenuation forecasting, where data-driven methods like RNNs~\cite{jacoby2025spatio} have been adopted to leverage historical data for rain mitigation and nowcasting applications~\cite{jacoby2021short,samad2020learning}, complementing traditional time-series approaches. In prior work, we utilized weather-aware capacity predictions for network routing \cite{kadota2022switching}. In contrast, in this work, we forecast future capacity \emph{distributions} and use uncertainty estimates to proactively mitigate capacity shortage risks. 

\section{Problem Formulation and Oracle Solution}
\label{sec:formulation}

Consider a mmWave network with star-like topology\footnote{To extend the \system framework to general network topologies, the problem formulation would need to account for the impact of accepting individual slices on the entire network, which includes adding flow conservation constraints to~\eqref{eq:ilp_obj}-\eqref{eq:ilp_vars}.} (similar to Fig.~\ref{fig.design1}, detailed in \cite[Appendix~\ref{app:network}]{ourArXiv}) supporting different 5G services including Ultra-Reliable and Low Latency Communications (URLLC), Enhanced Mobile Broadband (eMBB), Massive Machine-Type Communications (mMTC), and Best-Effort (BE). \emph{Users} generate traffic flows with different QoS requirements. \emph{Tenants} aggregate user flows and generate SRs associated with different 5G service types. The \textit{infrastructure provider} utilizes an admission control algorithm to decide whether to accept or reject arriving SRs to a particular mmWave link, aiming to maximize revenue.

\noindent\textbf{Link Capacity.} 
Time is divided into time-slots with index $t \in \{1, \ldots, \mathcal{H}\}$, where $\mathcal{H}$ is the time-horizon and the slot duration is $1$ minute. Capacity measurements, denoted by $\{C_t\}_{t=1}^{\mathcal{H}}$, fluctuate over time, potentially dropping to zero during severe rain fading events. We describe the measurements collected from real-world mmWave links in Sec.~\ref{sec:data_mesh}. 

\noindent\textbf{Slice Requests.} SRs are ordered chronologically with index $i\in\{1,\ldots,N\}$, where $N$ is the number of SRs generated within the $\mathcal{H}$ time-slots. In Sec.~\ref{sec:slice}, we describe the SR generation model, based on real-world network traces. SR~$i$ is characterized by service type (URLLC, eMBB, mMTC, or BE), throughput requirement $d_i$, duration $D_i$, sum reward $R_i= \rho_i d_i D_i$ for admitting SR~$i$, where $\rho_i$ is the price per unit throughput per slot (which varies for different service types), and a per slot penalty $P_{it}$ incurred when $d_i$ cannot be fulfilled in slot $t$. Let $t_i$ represent the slot when SR~$i$ is generated. Since several SRs can be generated in the same slot, it follows that $t_1 \leq t_2\leq\ldots\leq t_N$. Let the binary decision variable $z_i \in \{0,1\}$ indicates whether the algorithm admits SR~$i$. If admitted, $z_i=1$, then SR~$i$ becomes active during $t\in\{t_i,t_i+1,\ldots,t_i+D_i-1\}$. The indicator $k_{it}\in\{0,1\}$ indicates if SR~$i$ is active in slot $t$. It follows that $k_{it}\leq z_i,\forall i,t$.

\noindent\textbf{Underprovisioning and Penalty.} We consider admission control algorithms that do not admit new SRs when current admitted SRs are being underprovisioned. In slot $t$, underprovisioning occurs when $C_t<\sum_{i=1}^Nk_{it}d_i$. In this case, admission control algorithms attempt to allocate the available capacity $C_t$ to active SRs, aiming to minimize penalty. Let the decision variable $f_{it}\in[0,1]$ be the fraction of the requested throughput $d_i$ assigned to the active SR~$i$ in time slot $t$. The penalty for assigning $f_{it}<1$ in slot $t$ is a convex piecewise-linear function with $m_i$ segments defined as $P_{it}=\max_{k\in\{1,\ldots,m_i\}}\{a_{ik}(1-f_{it})+b_{ik}\}$ with penalty coefficients $a_{ik}\geq0$ and $b_{ik}\leq0$ that may vary for different 5G service types. Penalty coefficients are chosen such that $P_{it}=0$ when $f_{it}=1$ and $P_{it}$ increases sharply as $f_{it}$ approaches $0$ to encourage distributing the available capacity across all active SRs. In time slot $t$, admission control algorithms select $f_{it}$ to minimize $\sum_{i=1}^N P_{it}$ under the constraint $\sum_{i=1}^Nk_{it}d_if_{it}\leq C_t$. 

\noindent\textbf{Perfect-information MILP formulation (Oracle)} Assuming perfect knowledge of past, present, and future SRs and link capacity fluctuation, we design an Oracle admission control algorithm that maximizes the cumulative reward minus penalty. Specifically,  
we formulate 
a Mixed-Integer Linear Program (MILP) over the finite horizon $\mathcal{H}$ that 
serves as an idealized benchmark. 
The MILP is related to the classical Multidimensional Knapsack Problem, with two key distinctions. First, the knapsack capacity (i.e., the available link capacity $C_t$) changes over time. Second, items can be fractionally packed $f_{it}<1$ 
with a penalty $P_{it}>0$. The MILP formulation shown below jointly determines: SRs to admit $z_i,\forall i$, and (ii) fractions $f_{it},\forall i,t$, of the SRs throughput $d_i$ to allocate. Key notation used in this work is summarized in \cite[Appendix~\ref{app:notation}]{ourArXiv}. 
\begin{align}
\min \quad 
& \textstyle\sum_{t=1}^\mathcal{H} \textstyle\sum_{i=1}^N P_{it}
\;-\;
\textstyle\sum_{i=1}^N z_i R_i
\label{eq:ilp_obj} \\[6pt]
\text{s.t.} \quad
& z_i = 1 \;\rightarrow\;
P_{it} \ge a_{ik}(1 - f_{it}) + b_{ik},
\quad \forall i,t,k
\label{eq:ilp_pen} \\[6pt]
& \textstyle\sum_{i=1}^N k_{it} d_i f_{it} \le C_t,
\quad \forall t
\label{eq:ilp_capacity} \\[6pt]
& P_{it} \ge \delta \;\rightarrow\; z_{i'} = 0,
\quad
\forall i,t,\;
\forall i' \text{ such that } t_{i'} = t
\label{eq:ilp_non_admission} \\[6pt]
& f_{it} = 0, P_{it} = 0
\quad \forall i,t \text{ such that } k_{it}=0
\label{eq:ilp_f_out} \\[6pt]
& \delta > 0 ,\; 0 \le f_{it} \le 1,\;
z_i \in \{0,1\},\;
P_{it} \ge 0 
\label{eq:ilp_vars}
\end{align}
The objective function~\eqref{eq:ilp_obj} minimizes penalty minus reward, which is equivalent to maximizing revenue. 
Constraint~\eqref{eq:ilp_pen} uses an epigraph reformulation of the penalty function. 
Constraint~\eqref{eq:ilp_capacity} enforces the time-varying capacity limit. 
Constraint~\eqref{eq:ilp_non_admission} prevents the admission of new SRs while underprovisioning, i.e., when $P_{it} \ge \delta>0$. 
Constraints~\eqref{eq:ilp_pen} and~\eqref{eq:ilp_non_admission} can be implemented using standard big-$M$ linearization or indicator constraints.
Although the Oracle solution to \eqref{eq:ilp_obj}-\eqref{eq:ilp_vars} cannot be applied in practice, due to the assumption of perfect information, it provides a benchmark for our evaluation in Sec.~\ref{sec:evaluation}. Further, it guides the design of the \system framework that leverages short-term predictions of $C_t$ to inform its online admission control decisions. Next, we describe both datasets utilized in the design and evaluation of \system.



\section{Datasets: Link Attenuation and Slice Requests} \label{sec:data}

\subsection{mmWave Link Measurements} \label{sec:data_mesh}
\noindent\textbf{Dataset.} In this paper, we use networking logs collected by NYC Mesh~\cite{jacoby2025openmeshdata} containing Received Signal Level (RSL) measurements at one-minute resolution from 25 directional V-band (58--70 GHz) links with distances from hundreds of meters to several kilometers. 
RSL measures received power, in dBm, which is directly affected by rain-induced attenuation. 
Details about the mmWave network infrastructure (partially shown in Fig.~\ref{fig.design1}) and the dataset (that we made publicly available~\cite{jacoby2025openmeshdata}) can be found in~\cite[Appendix~\ref{app:data_availability}]{ourArXiv}. 

\begin{figure}[t]
\vspace{0.1em}
    \centering
    \includegraphics[width=\linewidth]{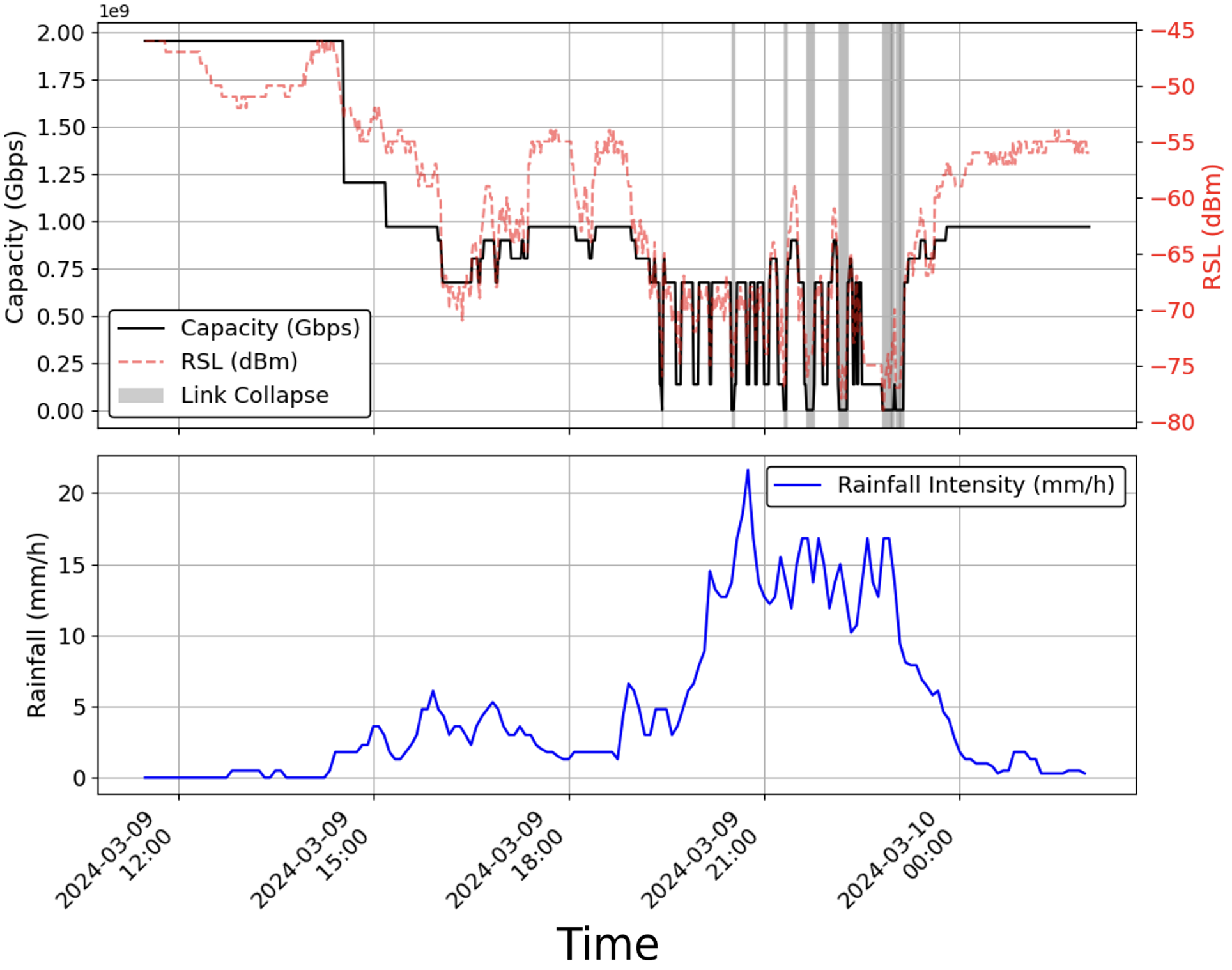}
\vspace{-0.2cm}
\caption{RSL and capacity fluctuation during rainfall event.}\label{fig:rain_event}
\vspace{-2em}
\end{figure}

\noindent\textbf{Weather-Induced Capacity Fluctuation.} 
Most prior work on slice admission control assumes static link capacity, whereas real-world networks can be sensitive to weather-induced attenuation. V-band links (shown in red in Fig.~\ref{fig.design1}) are highly susceptible to precipitation, which can cause severe absorption and scattering \cite{ITU, nandi2018study}, leading to sharp drops in RSL and link capacity $C_t$, posing a critical challenge for network operators. 
Fig.~\ref{fig:rain_event} showcases this impact on a representative 69 GHz link (March 09--10, 2024): while the signal remains stable during dry periods, heavy rain results in attenuation exceeding 30 dB, leading to temporary outages. To map RSL to discrete link capacity values, we adopt a standard hysteresis-based Adaptive Coding and Modulation (ACM)  approach, described in Section~\ref{sec:predictor} and detailed in  \cite[Appendix~\ref{app:ACM}]{ourArXiv}.

\subsection{Modeling Slice Requests from Network Traces}\label{sec:slice}

Standardized mechanisms for slice generation have not yet been established in the industry. We propose a SR generation pipeline based on network traces. 


\noindent\textbf{Dataset and 5G Services.} For over one month, we collected user traffic demands from 520 Columbia University residential units
in off-campus buildings. These residential buildings accommodate graduate students, postdocs, and faculty and their families. We anonymized privacy-sensitive fields, discarded personally identifiable information, and publicized the dataset~\cite{Yu2025Residential} with network traces. In this paper, we group packets with the same 5-tuple (i.e., source IP, destination IP, source port, destination port, transport protocol) as a flow. We associate the most recent DNS lookup to each flow. Based on the DNS domains, TLS Server Name Indications (SNI), port numbers, and IP addresses, we associate each flow with a specific 5G service according to the following mapping:
%
%
    (i)~\textbf{eMBB} service is designed for high-speed, high-capacity data transmission. We associate eMBB SRs with flows from video streaming and gaming services, including YouTube, Netflix, HBO, Hulu, Steam, and PlayStation; 
    (ii)~\textbf{URLLC} service is designed for ultra-reliable, low-latency communication services. 
    We associate URLLC SRs with flows from video conferencing and chatting applications, including Zoom, Snapchat, WhatsApp, and GoogleMessages; 
    (iii)~\textbf{mMTC} service caters to the connectivity needs of numerous IoT devices. 
    We did not identify IoT flows in our data; 
    (iv)~\textbf{BE} service is designed for non-critical communication. We associate BE SRs with flows from Spotify, Gmail, and Google Docs. 

\noindent\textbf{Slice Request Characterization.} For management simplicity, we assume that there is only a limited number of SR types, each type characterized by 5G service, duration $D_i$, throughput $d_i$, reward $R_i$, and penalty $P_{it}$. Rewards and penalties are defined by tenants and infrastructure providers based on the 5G service. In Sec.~\ref{sec:evaluation} we evaluate \system using different $R_i$ and $P_{it}$. We compute SR duration $D_i$ and throughput $d_i$ based on flow statistics. We consider three $D_i$ options: the 90th percentile of flow duration for each of the three 5G service categories. For each $D_i$, we consider four $d_i$ options: the 25th, 50th, 75th, and 95th percentiles of aggregated flow throughput. Throughput values are scaled up by a factor of 4× to ensure that demands are aligned with the available link capacity in our evaluation, and to anticipate increased bandwidth consumption due to the growing adoption of data-intensive services. 
In total, we generate 12 types of SRs. 


\noindent\textbf{Slice Request Generation.} With the above principles, we consider flows in chronological order and generate SRs as follows: For each incoming flow, if its throughput cannot be accommodated within active SRs, then a new SR is generated using the flow’s rounded-up throughput. When a flow lasts beyond its SR, the flow is treated as an incoming flow.

\section{System Design} \label{sec.method}
\system\ integrates an AttIRNN predictor with a two-tier SR admission controller. \system Workflow: 
(1)~Recent RSL measurements $\mathbf{X}_t$ are fed into the trained AttIRNN predictor.
(2)~The predictor outputs short-term capacity distributions. 
(3)~Admission control maximizes expected revenue using either: (i) Locally Optimal 
or (ii) Predictive Q-Learning. 

\subsection{AttIRNN Predictor} \label{sec:predictor}
AttIRNN is a multi-step probabilistic predictor of RSL values that leverages the significant temporal correlation of weather-induced fluctuations over short periods (see Fig.~\ref{fig:rain_event}). AttIRNN uses a short-term horizon of $H=5$ minutes. AttIRNN transforms the sequence of RSL predictions into capacity distributions that are input to admission control. 

\noindent\textbf{Prediction Problem Formulation.} Given the (static) link characteristics 
$\mathbf{x}_s \in \mathbb{R}^S$
(e.g., frequency and length~\cite{ITU}) and the $T$-step input history $\mathbf{X}_{t} = \{x_{t-T+1}, \dots, x_{t}\} \in \mathbb{R}^{T}$ which represents the $T$ most recent RSL measurements, AttIRNN learns a mapping
$F_{\Theta}: (\mathbf{X}_{t}, \mathbf{x}_s) \mapsto 
\hat{\mathbf{Y}}_{t}$,
where $\hat{\mathbf{Y}}_{t} = \{(\hat{\mu}_{t+1}, \hat{\sigma}_{t+1}^2), \dots, (\hat{\mu}_{t+H}, \hat{\sigma}_{t+H}^2)\} \in \mathbb{R}^{H \times 2} $ represents the $H$-step predicted mean–variance pairs with $\hat{\mu}_{t+h}$ estimating the future expected RSL $x_{t+h}$ and $\hat{\sigma}_{t+h}^2$ capturing the inherent uncertainty of the prediction at step $t+h$. 

\noindent\textbf{Learning Setup.}
The training set is built with a sliding window of length $T+H$, producing supervised pairs 
$(\mathbf{X}, \mathbf{Y})$ where 
$\mathbf{X} = \{x_1,\dots,x_T\}$ 
and 
$\mathbf{Y} = \{y_1,\dots,y_H\}$ with $y_h = x_{T+h}$. At inference, these map to time-indexed $\mathbf{X}_t$ and $\hat{\mathbf{Y}}_{t}$. 
Model parameters $\Theta$ are learned by minimizing the Gaussian negative log-likelihood (NLL)
\begin{equation}
\mathcal{L}(\Theta) = \frac{1}{2}
\sum_{(\mathbf{X}, \mathbf{Y}) \in \mathcal{D}} \sum_{h=1}^{H}
\left[
\frac{(y_h - \hat{\mu}_h)^2}{\hat{\sigma}_h^{2}}
+ \log \hat{\sigma}_h^{2}
\right],
\label{eq:nll}
\end{equation}
which drives accurate mean predictions and calibrated predictive uncertainty, assuming independent\footnote{While the errors are assumed independent and Gaussian per horizon step, the actual predictions are not assumed independent.} Gaussian prediction errors $y_h - \hat{\mu}_h \sim \mathcal{N}(0, \hat{\sigma}_h^{2})$ at each step $h$.

\begin{figure}[t]
\centering
\includegraphics[width=\columnwidth]{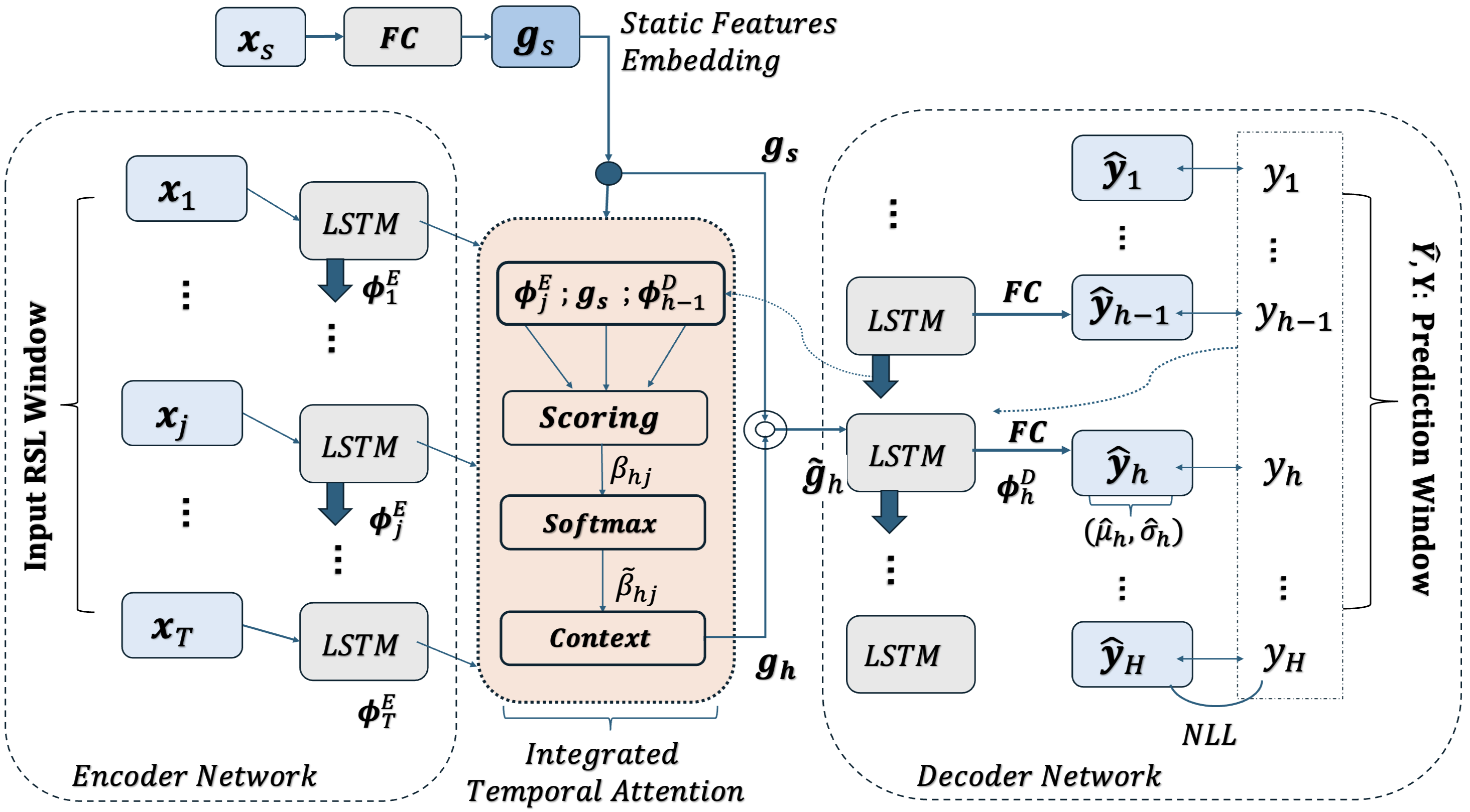}
\caption{AttIRNN. Seq2Seq-based predictions with temporal attention for \(H\)-step prediction using \(T\)-step history.}
\label{fig.attIRNN}
\vspace{-2em}
\end{figure}

\noindent\textbf{AttIRNN Architecture.} 
Fig.~\ref{fig.attIRNN} depicts the AttIRNN architecture centered around 
an attention-augmented encoder-decoder framework~\cite{cho2014learning}.
The encoder processes the input sequence $\mathbf{X}$ via LSTM cells~\cite{hochreiter1997long} to generate hidden states $\{\boldsymbol{\phi}_j^E\}_{j=1}^T$. 
Given a link's static features $\mathbf{x}_s \in \mathbb{R}^S$, AttIRNN computes an embedding $\mathbf{g}_s$ via a fully-connected (FC) layer. This embedding enables link-specific adaptation while leveraging shared patterns learned across multiple links.
At each decoder step $h$, attention aligns the previous decoder state $\boldsymbol{\phi}_{h-1}^D$, static context $\mathbf{g}_s$, and encoder states via:
\begin{align}
\beta_{hj} &= \mathbf{v}_a^\top \tanh\big(\mathbf{W}_a[\boldsymbol{\phi}_{h-1}^D;\, \mathbf{g}_s;\, \boldsymbol{\phi}_j^E] + \mathbf{b}_a\big), \\
\tilde{\beta}_{hj} &= \exp(\beta_{hj})/\textstyle\sum_{j'=1}^T \exp(\beta_{hj'}),
\end{align}
yielding context $\mathbf{g}_h = \sum_{j=1}^T \tilde{\beta}_{hj}\boldsymbol{\phi}_j^E$ and augmented context $\tilde{\mathbf{g}}_h = [\mathbf{g}_h;\, \mathbf{g}_s]$.
The decoder then autoregressively generates predictions through states $\{\boldsymbol{\phi}_h^D\}_{h=1}^H$:
\begin{align} \label{eq:attirnn_decoder}
\boldsymbol{\phi}_h^D &= \text{LSTM}\big(y_{h-1},\, [\boldsymbol{\phi}_{h-1}^D,\, \tilde{\mathbf{g}}_h]\big), \\
\hat{\mathbf{y}}_h &= \mathbf{W}_o[\boldsymbol{\phi}_h^D;\, \tilde{\mathbf{g}}_h] + \mathbf{b}_o,
\end{align}
where $\hat{\mathbf{y}}_h = (\hat{\mu}_h, \hat{\sigma}_h^2)$ with positive $\hat{\sigma}_h^2$ via softplus. Here $y_{h-1}$ is ground truth at training (teacher forcing) and previous prediction $\hat{\mu}_{h-1}$ at inference.

\noindent\textbf{Discrete Capacity Distributions.} 
During regular operations, mmWave links use ACM to map their current RSL to a discrete link capacity. We consider the standard hysteresis-based ACM detailed in \cite[Appendix~\ref{app:ACM}]{ourArXiv} with capacity levels $C_l$, with $C_0<C_1<\cdots<C_{L-1}$, and associated RSL hysteresis thresholds $C_l^{\downarrow}$ (to transition from $C_{l}$ to $C_{l-1}$) and $C_l^{\uparrow}$ (to transition from $C_{l}$ to $C_{l+1}$). 

At slot $t$, AttIRNN uses the current (known) capacity $C_{l'}$ and the $H$-step mean-variance pairs $\hat{\mathbf{Y}}_{t} = \{(\hat{\mu}_{t+h}, \hat{\sigma}_{t+h}^2)\}_{h=1}^H$ to generate probability mass functions $\tilde{\mathbf{p}}_{t+h} = \{\tilde{p}_{l,h}\}_{l=0}^{L-1}$ over the capacity levels $C_l\in\{0,\ldots,L-1\}$ and over the short-term horizon $h\in\{1,\ldots,H\}$, where $\tilde{p}_{l,h}\in[0,1]$ represents the probability that the link capacity at time slot $t+h$ is $C_l$. To obtain $\tilde{p}_{l,h}$, we use the assumption of independent Gaussian prediction errors to compute the $h$-step transition probabilities from the current level $C_{l'}$ to $C_l$ given by
\begin{equation}
\tilde{p}_{h}(l' \rightarrow l)=
\Phi\!\Bigl(\smash{\tfrac{\max(C_l^{\uparrow},C_{l'}^{\downarrow})-\hat\mu_{t+h}}{\hat\sigma_{t+h}}}\Bigr)-
\Phi\!\Bigl(\smash{\tfrac{\min(C_l^{\downarrow},C_{l'}^{\uparrow})-\hat\mu_{t+h}}{\hat\sigma_{t+h}}}\Bigr)
\label{eq:cap_transition}
\end{equation}
where $\Phi(\cdot)$ is the standard normal CDF. 
Then, assuming a first-order Markov model, 
the short-term capacity distribution $\tilde{\mathbf{p}}_{t+h},\forall h\in\{1,\ldots,H\}$ can be computed iteratively as
\begin{align}\label{eq:markov_recursion}
\tilde{p}_{l,h} &= \textstyle\sum_{j=0}^{L-1} \tilde{p}_h(j\to l)\,\tilde{p}_{j,h-1}, \quad h=2,\dots,H,
\end{align}
initialized by $\tilde{p}_{l,1} = \tilde{p}_1(l'\to l)$. These short-term capacity distributions that account for prediction uncertainty $\hat{\sigma}_h^{2}$ are input to slice admission control algorithms discussed next. 

\subsection{Slice Admission Control Algorithms} \label{sec.optimizer}
We present three slice admission algorithms: (i) a Rate Control mechanism for SR underprovisioning, (ii) a Locally Optimal algorithm that leverages capacity distributions for revenue maximization, and (iii) a Predictive Q-Learning algorithm that incorporates capacity forecasts into RL. 

\noindent\textbf{Rate Control for SR Underprovisioning in Slot $\mathbf{t}$.} Denote by $\mathcal{N}_t=\{i\in\{1,\ldots,N\}|k_{it}=1\}$ the set of active SRs at the beginning of slot~$t$. In slot $t$, underprovisioning occurs when $C_t<\sum_{i\in\mathcal{N}_t}d_i$. In this case, instead of admitting new SRs, admission control algorithms select 
fractions $f_{it},\forall i\in\mathcal{N}_t$ of the active SRs throughput request $d_i$ aiming to minimize the QoS-violation penalty in slot $t$ subject to the link capacity constraints, as described by the linear problem below 
\begin{align}
\min \quad 
& \textstyle\sum_{i\in\mathcal{N}_t} P_{it}
  \label{eq:lp_obj} \\
\text{s.t.} \; & P_{it} \geq a_{ik}\bigl(1 - f_{it}\bigr) + b_{ik}, \forall \, k, \, i \in \mathcal{N}_t \label{eq:lp_pen} \\
& \textstyle\sum_{i\in\mathcal{N}_t} d_i f_{it} \le C_t
\label{eq:lp_capacity} \\
& 0 \le f_{it} \le 1,\quad P_{it} \ge 0
\label{eq:lp_vars}
\end{align}
This rate control mechanism is integrated into both the Locally Optimal and Predictive Q-learning algorithms described next.


\noindent\textbf{Locally Optimal (LO).} In each slot~$t$, LO considers three inputs: (i) the set of active SRs $\mathcal{N}_{t}$; (ii) the batch of arriving SRs, denoted by $\mathcal{A}_t=\{i\in\{1,\ldots,N\}|t_i=t\}$; and (iii) the capacity distributions $\tilde{\mathbf{p}}_{t+h}$. The goal of LO is to admit the subset of arriving SRs, $\mathcal{A}^*_t\subseteq\mathcal{A}_t$, that maximizes the expected short-term revenue $\mathbb{E}\left[\sum_{i\in\mathcal{A}^*_t}R_i-\sum_{\tau=t}^{t+H} \sum_{i\in\{\mathcal{N}_{t}\cup\mathcal{A}^*_t\}} P_{i\tau}\right]$. 
In slot $t$, there are $2^{|\mathcal{A}_t|}$ SR subsets that can be admitted (each subset is associated with a reward $\sum_{i}R_{i}$) and $L^H$ possible sequence of capacity levels $C_l$ over the short-term horizon (each sequence is associated with a probability $\prod_{h=1}^H\tilde{p}_{l,h}$). Each of the $2^{|\mathcal{A}_t|}L^H$ combinations is associated with a penalty $\sum_{\tau}\sum_{i}P_{i\tau}$ that may affect the expected short-term revenue. 

To reduce complexity, a key insight is that for each SR subset, the penalty $\sum_{i}P_{i\tau}$ obtained from \eqref{eq:lp_obj}-\eqref{eq:lp_vars} only depends\footnote{Here, we make the pessimistic assumption that no SR will expire within the short-term horizon $H$.} on the capacity level $C_{\tau}\in\{C_0,\ldots,C_{L-1}\}$ and not on $\tau$. Hence, to obtain $\sum_{i}P_{i\tau},\forall \tau\in\{t,\ldots,t+H\}$, we only need to solve \eqref{eq:lp_obj}-\eqref{eq:lp_vars} $L$ times, instead of $L^H$ times. When $|\mathcal{A}_t|$ is small, LO can easily compute the  $2^{|\mathcal{A}_t|}$ possible rewards and corresponding expected short-term revenues to obtain the optimal SR subset $\mathcal{A}^*_t\subseteq\mathcal{A}_t$. When $|\mathcal{A}_t|$ is large (in our evaluation it can be $\sim100$ SRs), LO finds an approximate solution by sorting the set $\mathcal{A}_t$ in decreasing order of rewards $R_i$ and adding SRs to the admitted set $\mathcal{A}^*_t$, one at a time, if their addition increases the short-term expected revenue, thereby reducing complexity from $2^{|\mathcal{A}_t|}$ to ${|\mathcal{A}_t|}$. 

In contrast with Oracle, the LO algorithm does not assume knowledge of future SR batches. 
As a result, LO may admit low-reward SRs in slot $t$ that prevent the admission of high-reward SRs in the near future, $t+h$, leading to suboptimal overall performance. Predictive Q-learning addresses this limitation by interacting with the environment to learn the ``quality'' of taking specific actions.

\noindent\textbf{Predictive Q-Learning (PQL).} PQL extends standard Q-learning~\cite{watkins1992qlearning} by explicitly incorporating predictions of future link capacity into the SR admission process. 
In each slot~$t$, PQL considers the same three inputs as LO. PQL sorts arriving SRs within $\mathcal{A}_t$ in decreasing order of rewards $R_i$. Let $i(j)\in\{\mbox{sort}(\mathcal{A}_1),\mbox{sort}(\mathcal{A}_2),\ldots,\mbox{sort}(\mathcal{A}_{\mathcal{H}})\},\forall j\in\{1,2,\ldots,N\}$ represent the re-ordered SR index $i$ sorted in decreasing value of rewards $R_i$ within each set $\mathcal{A}_t$. PQL evaluates the addition of SRs to the admitted set $\mathcal{A}^*_t$, one at a time, using an $\epsilon$-greedy policy in which PQL selects a random action $a_{i(j)}\in\{0,1\}$ (to reject or admit SR $i(j)$) with probability $\epsilon$ (exploration) or PQL selects actions according to its Q-table $Q(s_{i(j)},a_{i(j)})$ with probability $1-\epsilon$ (exploitation). The exploration rate \( \epsilon \) decays over iterations $j$. 
After selecting action $a_{i(j)}$, PQL receives reward $r(s_{i(j)},a_{i(j)})$, updates its state from $s_{i(j)}$ to $s_{i(j+1)}$, and considers the admission of SR $i(j+1)$. 

The Q-table $Q(s_{i(j)},a_{i(j)})$ estimates the expected long-term revenue of taking action $a_{i(j)}\in\{0,1\}$ in state $s_{i(j)} = (\mathbf{n}, \mathbf{A}_{\text{new}}, c_f)$, where $\mathbf{n} = (n_1, \ldots, n_{12})$ records the number of active SRs of each type (as defined in Section~\ref{sec:slice}), $\mathbf{A}_{\text{new}}$ is a one-hot encoding vector that indicates the type of SR $i(j)$, and $c_f$ represents the number of future time slots, within the short-time horizon $H$, in which capacity constraints $C_t$ are satisfied if SR $i(j)$ is admitted. 
If SR $i(j)$ is rejected, the reward $r(s_{i(j)}, a_{i(j)}=0)=0$. If SR $i(j)$ is admitted, the reward $r(s_{i(j)}, a_{i(j)}=1)=R_{i(j)}-R_{i(j)}(1-c_f/H)\lambda$, where $\lambda>0$ is a risk-penalty scaling coefficient that controls the cost of anticipated capacity violations and $(1-c_f/H)$ approximates the probability of capacity violation over the short-term horizon. Penalty scaling with $(1-c_f/H)$ allows PQL to account for the impact of current admission decisions on future underprovisioning. 

The Q-table is updated iteratively according to 
\begin{equation}
\begin{aligned}
Q_{new}(s_{i(j)}, a_{i(j)}) 
= Q(s_{i(j)}, a_{i(j)}) + \alpha_j \Bigl[r(s_{i(j)}, a_{i(j)})+\\
 \max_{a'\in \{0,1\}} Q\bigl(s_{i(j+1)}, a'\bigr) 
         \;-\; Q(s_{i(j)}, a_{i(j)})\Bigr].
\end{aligned}
\label{eq:q_update}
\end{equation}
where \( \alpha_j \) is the learning rate at iteration \( j\in\{1,\ldots,N\} \), determined by the occurrence count \( o(s_{i(j)}, a_{i(j)}) \) of the state-action pair \( (s_{i(j)}, a_{i(j)}) \). 
The learning rate is updated as $\alpha_j = 0.5/o(s_{i(j)}, a_{i(j)})$. This choice of \( \alpha_j \) ensures that the learning rate decreases over time as the agent becomes more experienced with the state-action pair, which is crucial for the convergence of the Q-learning algorithm. The discount factor is set to 1 to maximize long-term utility. 
Through iterative updates, PQL gradually learn SR admission policies that are robust to future capacity fluctuations and converges to a policy $Q^*(s,a)$ that maximizes the long-term expected revenue. The detailed description of PQL is given in Algorithm~\ref{alg:predictive_ql}.

{
\setlength{\textfloatsep}{1pt}
\begin{algorithm}[t]
\caption{Predictive Q-Learning (PQL) for SR Admission}
\label{alg:predictive_ql}
\scriptsize
\begin{algorithmic}[1]
\State \textbf{Initialize} Q-table $Q(s,a)\gets 0$, $\forall (s,a)$; occurrence counts $o(s,a)\gets 0$, $\forall (s,a)$
\State \textbf{Initialize} exploration parameters $\epsilon$, $\text{decay\_rate}$, $\text{min\_epsilon}$
\For{each time slot $t$}
    \State \textbf{(Optional)} If current allocations violate $C_t$, \textbf{Rate Control}, otherwise \textbf{Continue}
    \State Let $\mathcal A_t$ be the set of SRs arriving at slot $t$
    \If{$\mathcal A_t \neq \emptyset$}
        \State \textbf{Sort} $\mathcal A_t$ in decreasing order of rewards 
        \For{each SR $i(j)$}
            \State \textbf{Compute} predicted future capacity violations \( c_f \) in next 5 minutes 
            \State \textbf{Observe} state $s_{i(j)} \gets (\mathbf n,\mathbf A_{\text{new}},c_f)$
            \State \textbf{Use} \( \epsilon \)-greedy policy to select action \( a_{i(j)} \)
            \State \textbf{Execute} $a_{i(j)}$ (admit if $1$; reject if $0$) 
            \State \textbf{Compute} reward
            \State \hspace{1em} $r(s_{i(j)},a_{i(j)}) \gets 0$ if $a_{i(j)}=0$
            \State \hspace{1em} else $r(s_{i(j)},a_{i(j)}) \gets R_{i(j)} - R_{i(j)}(1-c_f/H)\lambda$
            \State \textbf{Observe} next state \( s_{i(j+1)} \)
            \State \textbf{Increment} occurrence count \( o(s_{i(j)}, a_{i(j)}) \leftarrow o(s_{i(j)}, a_{i(j)}) + 1 \)
            \State \textbf{Update} learning rate \( \alpha_t = 0.5/o(s_{i(j)}, a_{i(j)}) \)
            \State \textbf{Update} Q-value \( Q(s_{i(j)}, a_{i(j)}) \) using \eqref{eq:q_update}
        \EndFor
        \State \textbf{Decay} exploration rate: $\epsilon \leftarrow \max(\epsilon\cdot \text{decay\_rate},\ \text{min\_epsilon})$
    \EndIf
\EndFor
\end{algorithmic}
\end{algorithm}
}

\section{Evaluation}\label{sec:evaluation}
In this section, We evaluate AttIRNN prediction performance and the SR admission control performance. 
Using both real-world datasets described in Sec.~\ref{sec:data}, we evaluate \system using: (1) up to $30$ hours of SR generation; and (2) up to $34$ hours of measurements from six V-band links, spanning three precipitation events (March–April 2024), with rainfall intensities ranging from 5.45 to 45.7\,mm/h. To characterize the volatility of each simulated scenario, we compute its Coefficient of Variation ($\mathrm{CV}$), defined as the ratio of the standard deviation to the mean of the RSL measurements over 1-hour intervals. Higher CV indicates more volatile links, often due to intense rainfall.


In evaluating AttIRNN, we compare prediction baselines and architectural variations, with results averaged across all links and grouped by prediction horizon and CV. In evaluating the SR admission control algorithms, each simulated scenario consists of 1 hour of RSL measurements combined with 1 hour of SR generation. The volatility of the scenario is computed using CV. To characterize the maximum underprovisioning time fraction of each simulated scenario, we consider a fictional, non-admissible SR algorithm that admits all SRs and we compute the fraction of time slots the simulated scenario violates QoS. 
%
Details on SR generation and RSL measurements are provided in \cite[Appendix~\ref{app:data_availability}]{ourArXiv}.



\begin{figure}[t]
    \centering
\vspace{-1em}
    \subfloat[RSL Prediction RMSE\label{fig:rmse}]{%
        \includegraphics[width=0.49\columnwidth,height = 30mm]{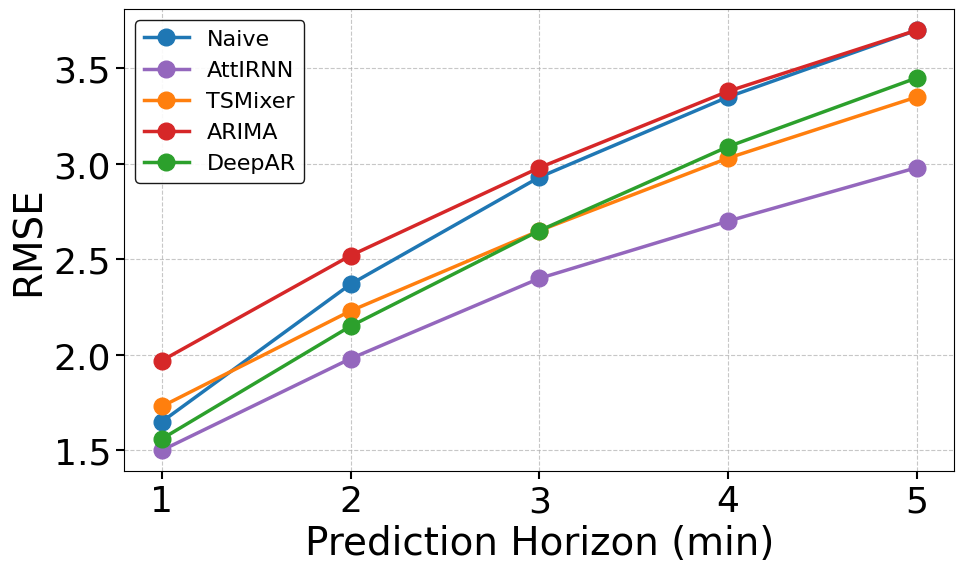}%
    }%
    \hfill
    \subfloat[AttIRNN Ablation Study\label{fig:ablation}]{%
        \includegraphics[width=0.49\columnwidth,height = 30mm]{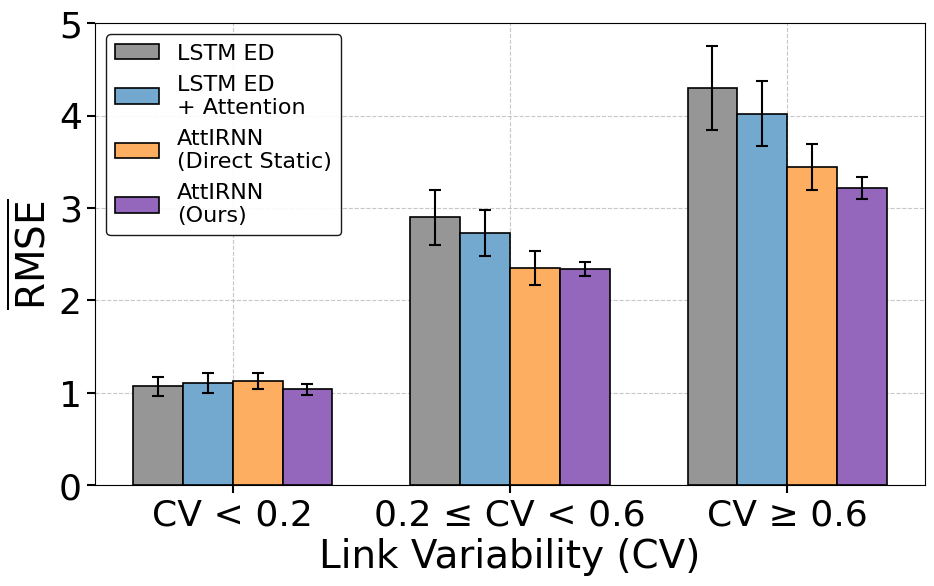}%
    }
        
    \subfloat[RSL Prediction Error Percentile\label{fig:quantiles}]{%
        \includegraphics[width=0.49\columnwidth,height = 30mm]{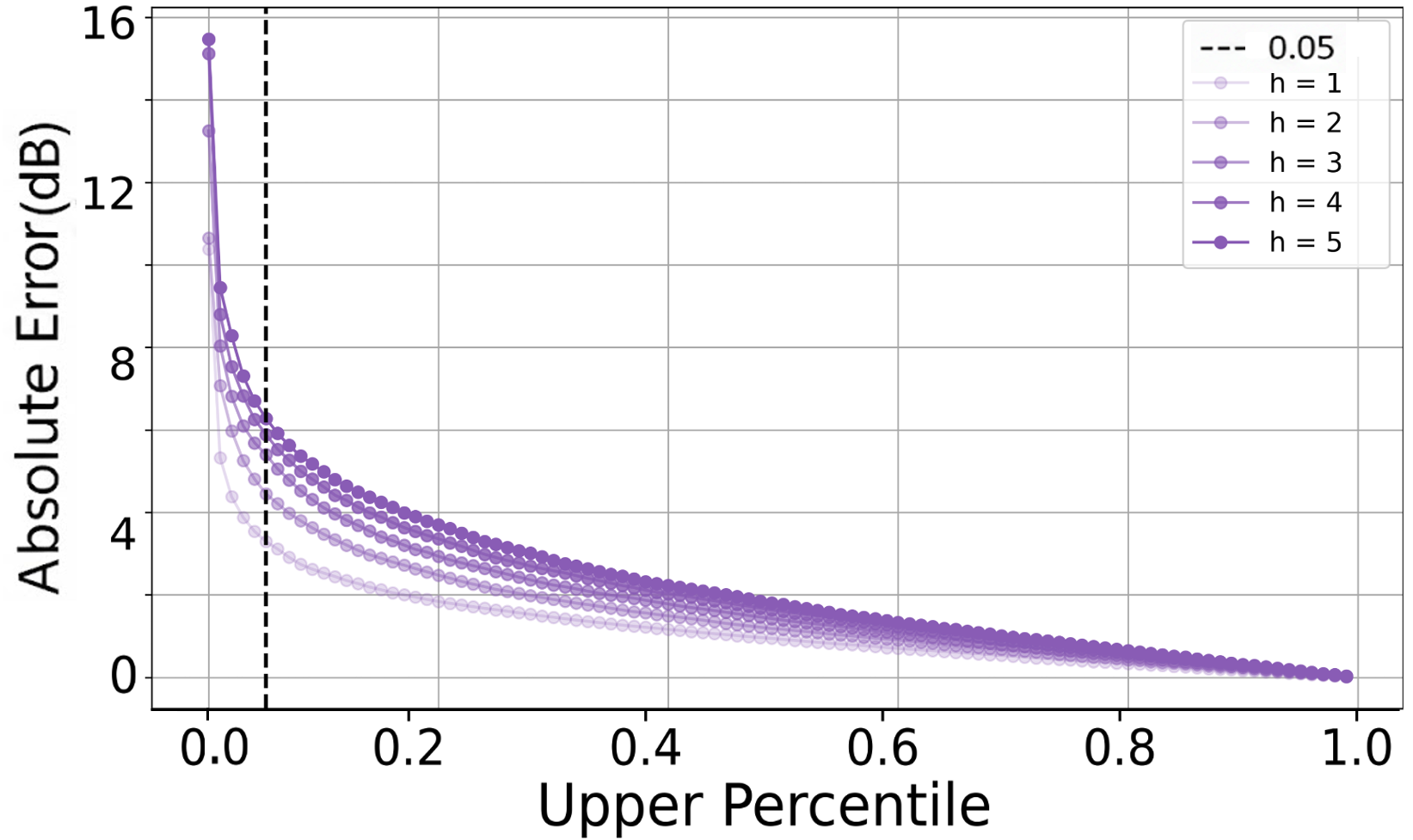}
    }%
    \hfill
    \subfloat[Future Capacity Prob. Distribution\label{fig:confidence}]{%
        \includegraphics[width=0.49\columnwidth,height = 30mm]{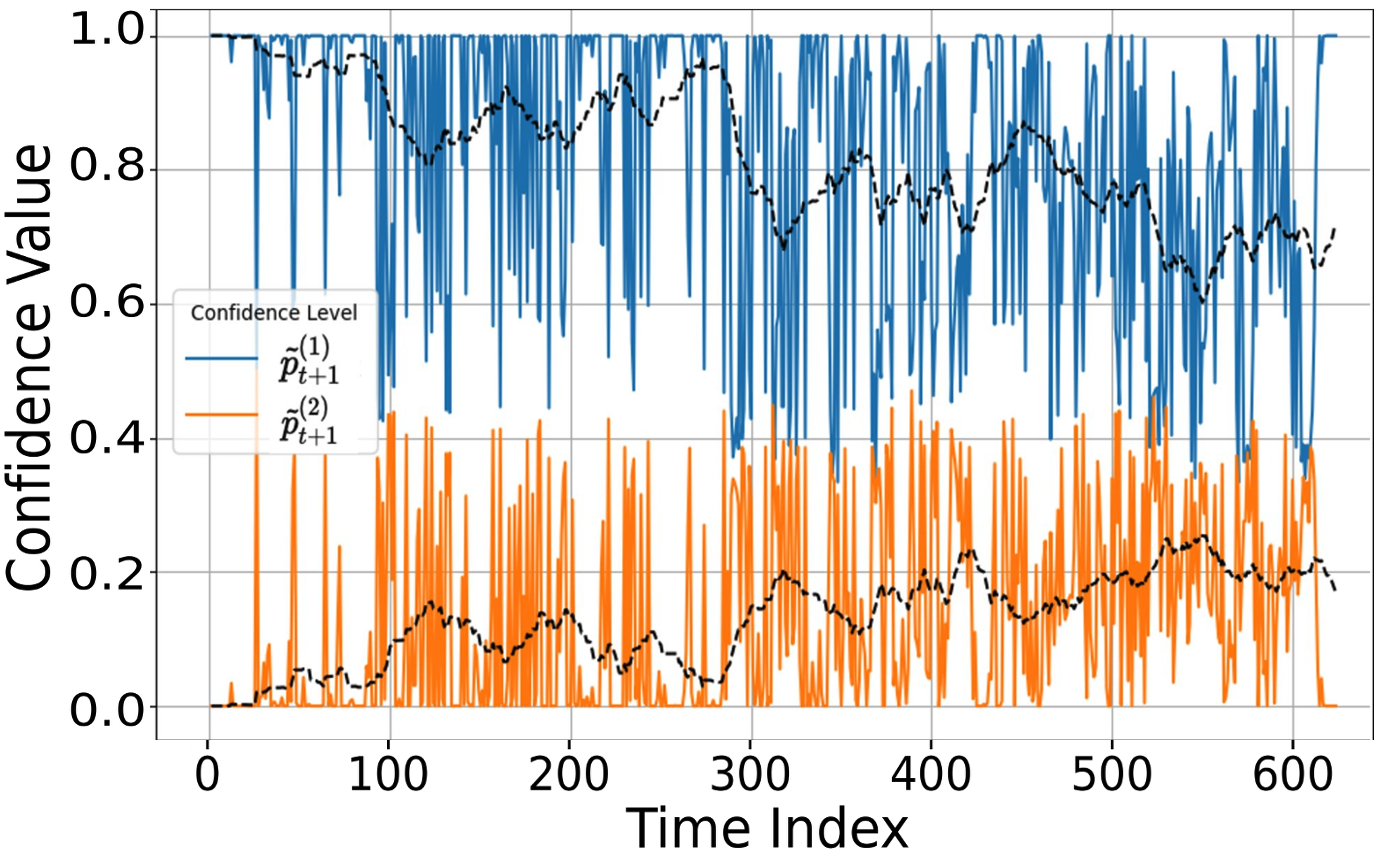}%
    }    
    \caption{(a) RSL Prediction RMSE for different SR algorithms as the short-term horizon increases; (b) Comparison of AttIRNN with LSTM encoder-decoder and temporal attention variants for different CV regimes; (c) RSL Prediction Error Percentile for as the short-term horizon increases; and (d) Evolution of the predicted probability distribution of link capacities over time.}
    \label{fig:pred_metrics}
    \vspace{-2em}
\end{figure}

\subsection{AttIRNN prediction performance} \label{sec:pred.evaluation}
We compare AttIRNN against established statistical and deep learning models, including: 
(i) Naive approach~\cite{hyndman2006another} which predicts $\hat{x}_{t+h} = x_t,\forall h,t$, being most effective in stable conditions; (ii) ARIMA~\cite{hyndman2008automatic}, which is a well-known baseline; (iii) TSMixer~\cite{chen2023tsmixer}, which is a recent MLP architecture that uses a Mean Squared Error (MSE) loss function; and (iv) DeepAR~\cite{salinas2020deepar}, which utilizes static features and a NLL loss~\eqref{eq:nll} for probabilistic forecasting. For AttIRNN, we use $H = 5$ horizon and $T = 15$ lookback. The static feature vector $\mathbf{x}_s$ includes link length, frequency, antenna type, and link ID to account for physical link characteristics that influence signal attenuation patterns. The AttIRNN architecture uses single-layer LSTMs with hidden dimension 64 for encoder and decoder, with link embeddings of dimension 8. 

We evaluate the different prediction models (Fig.~\ref{fig:rmse}) using Root Mean Square Error (RMSE) for prediction horizon $h$: $\text{RMSE}_h = \sqrt{\frac{1}{N}\sum_{t} (x_{t+h} - \hat{\mu}_{t+h})^2}$. 
%

\noindent\textbf{Training, Validation, and Evaluation Datasets.} We split the data from the 25 V-band links, collected within November 2023–April 2024, into: (i) training set, 19 links within November 2023–March 2024; (ii) validation set, 6 held-out links within November 2023–March 2024; and (iii) evaluation set, on the same 6 held-out links but within March–April 2024. All models were optimized via grid search during the training phase. 
%
Training uses Adam optimizer, batch size 128, up to 120 epochs with 10 epoch patience early stopping. Models were implemented in PyTorch on an NVIDIA RTX 2080 Ti. See \cite[Appendix~\ref{app:prediction_implement}]{ourArXiv} for details.



\noindent\textbf{RSL Prediction RMSE.} Fig.~\ref{fig:rmse} shows that AttIRNN consistently achieves the lowest RMSE for all prediction horizons. While most models perform comparably at $h=1$, statistical baselines (Naive, ARIMA) degrade rapidly, exceeding 3.5~dB at $h=5$. Fig.~\ref{fig:ablation} displays the mean RMSE across CV regimes, with error bars representing the standard deviation over 10 runs. The Direct-Static variant isolates metadata impact by feeding the static embedding $\mathbf{g}_s$ directly into the decoder, effectively bypassing the Integrated Temporal Attention layer. This configuration confirms that physical link characteristics are primary drivers of accuracy, particularly at high link variability where $\text{CV} \ge 0.6$. While this outperforms the LSTM ED and LSTM + Attention baselines that lack explicit static integration, our full AttIRNN (Ours) achieves the lowest RMSE and demonstrates stable training across all regimes. These results confirm that static integration provides the vital link-aware context while integrated attention captures dynamic temporal fluctuations to further refine the prediction.

\noindent\textbf{Prediction Error Percentile.} Fig.~\ref{fig:quantiles} shows the absolute RSL prediction error percentile for $h\in\{1,\ldots,H\}$. The dashed vertical line indicates that 95\% of all predictions exhibit an absolute error of less than 4~dB at $h=1$ and 6~dB at $h=5$. The rarity of large prediction errors ensures reliable capacity forecasting during typical rain-induced attenuation patterns.

\noindent\textbf{Future Capacity Probability Distribution.} Fig.~\ref{fig:confidence} shows the evolution of the two most probable values of the capacity probabilities $\tilde{p}_{l,h=1}$ over $l\in\{0,1,\ldots,L-1\}$, denoted as $\tilde{p}^{(1)}_{t+1}$ and $\tilde{p}^{(2)}_{t+1}$. 
Periods where $\tilde{p}^{(1)}$ drops coincide with rises in $\tilde{p}^{(2)}$, demonstrating the expected shifting of probability mass between discrete capacity levels as link conditions deteriorate during the rain event. This behavior confirms the model's ability to capture dynamic uncertainty across discrete capacity values rather than producing overconfident point estimates.


\subsection{SR Admission Control Performance}\label{sec:aware.evaluation}

We compare the following SR algorithms: 
(i)~Random, which admits each SR with probability $0.5$; 
(ii)~Naive Greedy, which admits SRs whenever capacity allows; 
(iii)~Locally Optimal, described in \ref{sec.optimizer}. To reduce complexity, we consider LO that finds an approximate solution by admitting arriving SRs one at a time, in decreasing order of rewards; 
(iv)~Naive Q-learning~\cite{bega2017optimising}, which assumes static link capacity $C_t=C,\forall t$; 
(v)~Predictive Q-learning, described in Sec.~\ref{sec.optimizer}, with $\lambda=0.5$. Naive and Predictive Q-learning employ lightweight table-based updates and converge within 100k training steps. These Q-learning algorithms have an average runtime of 6 ms per-time-slot when handling arriving batches of 100 SRs, making them suitable for real-time deployment. 
(vi)~DeepNetSlice~\cite{DNS} which is a Deep Reinforcement Learning model that employs an actor–critic architecture with Graph Convolutional Networks to capture dynamic capacity states, allowing the agent to optimize long-term SR admission decisions under fluctuating capacity. However, DeepNetSlice does not leverage capacity predictions for SR admission, what limits its ability to mitigate underprovisioning penalties.
(vii)~Oracle benchmark, described in Sec.~\ref{sec:formulation}. 
When underprovisioning occurs, all SR admission control approaches use Rate Control \eqref{eq:lp_obj}-\eqref{eq:lp_vars} to minimize their penalty. 
For fairness, all methods are evaluated using the same reward and penalty function.

\noindent\textbf{SR Generation and Underprovisioning Penalty.} Simulations include the 12 types of SRs described in Sec.~\ref{sec:slice}, four per 5G service. The price per throughput unit per slot is $\rho_i\in\{2.5,5,10\}$ for BE, eMBB, and URLLC SRs, respectively. The piecewise-linear penalty functions have $m_i=2$ segments each, with slope and intercept of the segments determined by the equations $a_{ik}(1-f_{it})+b_{ik}$. 
The coefficients are in Table~\ref{tab:penalty_coeffs}. 
Additional details about the penalty coefficients utilized in our evaluation can be found in~\cite[Appendix~\ref{app:penalty}]{ourArXiv}.
\begin{table}[t]
\centering
\caption{Coefficients of the Penalty Functions used in Fig.~\ref{fig:combined_revenue}.}
\label{tab:penalty_coeffs}
\begin{tabular}{rcccc}
\toprule
& \multicolumn{2}{c}{Segment 1} & \multicolumn{2}{c}{Segment 2} \\
\midrule
5G Service & Slope & Intercept & Slope & Intercept \\
\midrule
URLLC & 6 & -2 & 2 & 0 \\
eMBB & 3 & -1 & 1 & 0 \\
BE & 1.5 & -0.5 & 0.5 & 0 \\
\bottomrule
\end{tabular}
\vspace{-2em}
\end{table}


\noindent\textbf{Q-Learning Training and Evaluation.} For the SR admission control experiments, we train Q-learning using all available SR generation and link-capacity data described in Sec.~\ref{sec:data}. Specifically, Q-learning updates its Q-table online and converges over the entire collection of SR generation and mmWave link-capacity traces. After convergence, we freeze the learned Q-table and evaluate the resulting admission policy on each individual simulated scenario, where each scenario consists of one hour of SR generated from residential traffic traces combined with one hour of measured link capacity from a V-band mmWave link. All baseline admission control algorithms are evaluated on the same set of scenarios, ensuring a fair comparison across methods.

\begin{figure}[t]
    \centering
    
    \subfloat[Cumulative Revenue]{%
        \includegraphics[width=0.49\columnwidth]{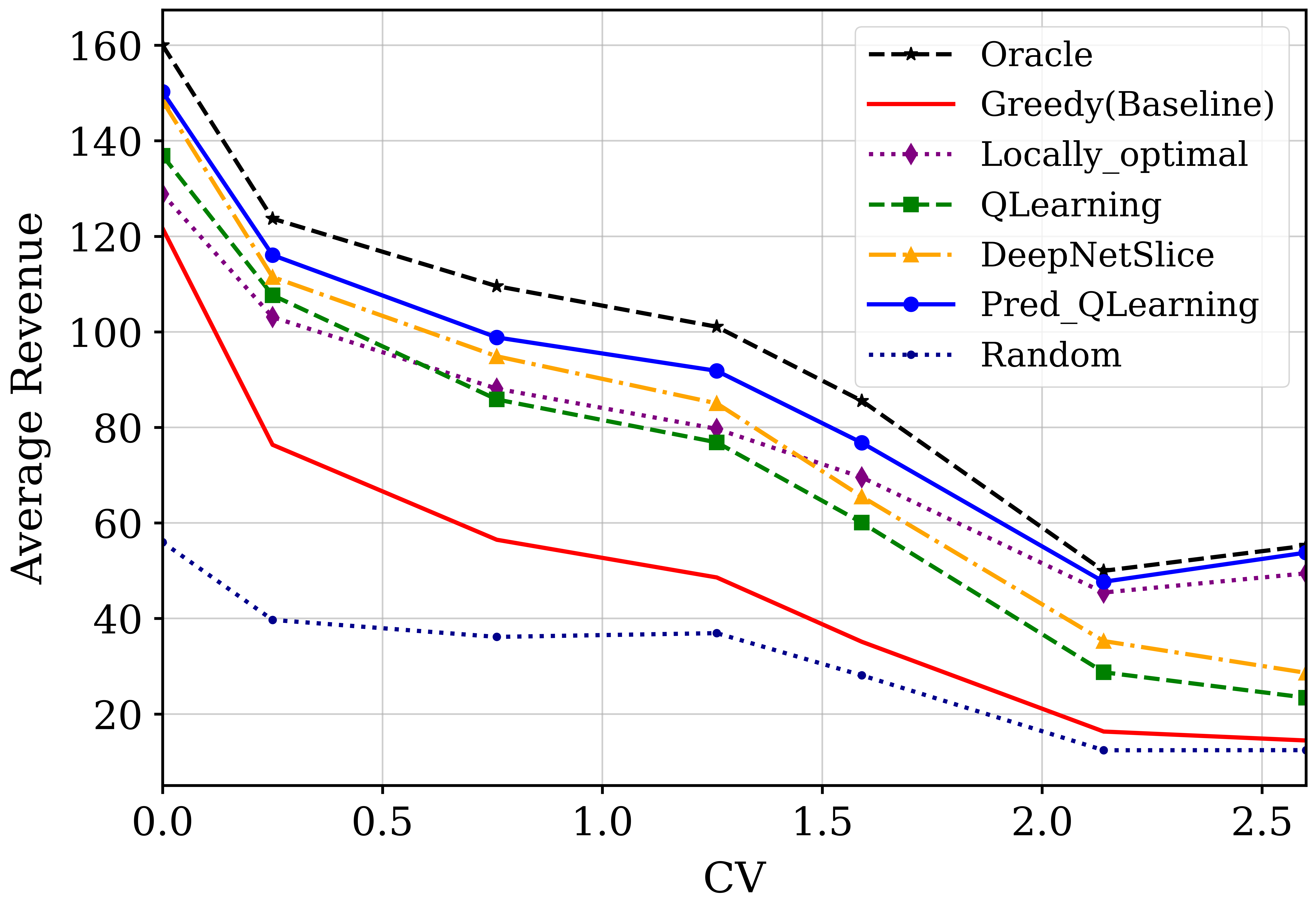}\vspace{-2em}%
        \label{subfig:average}
    }%
    \subfloat[Normalized Revenue]{
        \includegraphics[width=0.49\columnwidth]{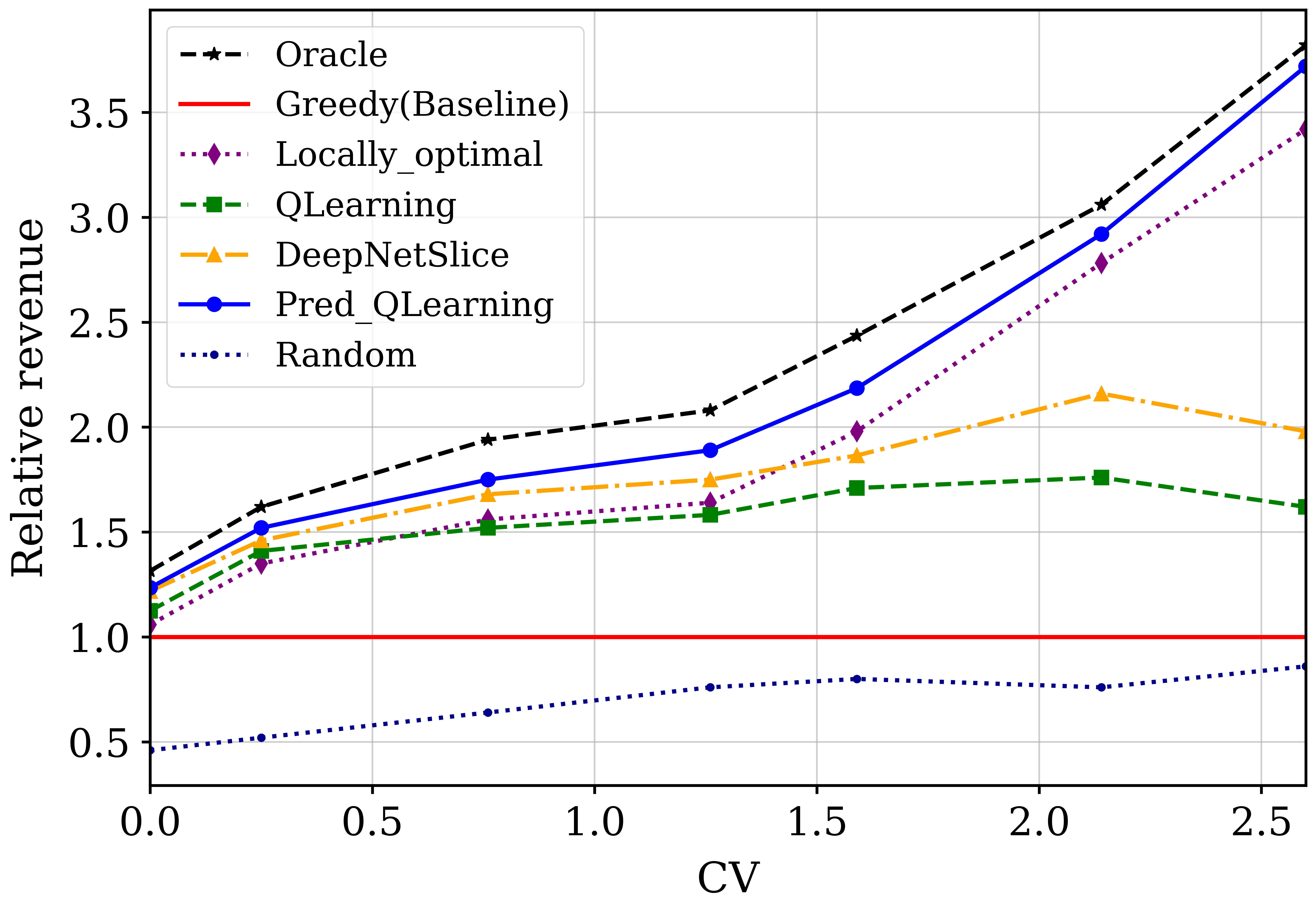}\vspace{-2em}%
        \label{subfig:random}
    }
     \hfill
    \subfloat[$\%$ of SRs with Negative Revenue]{%
        \includegraphics[width=0.49\columnwidth]{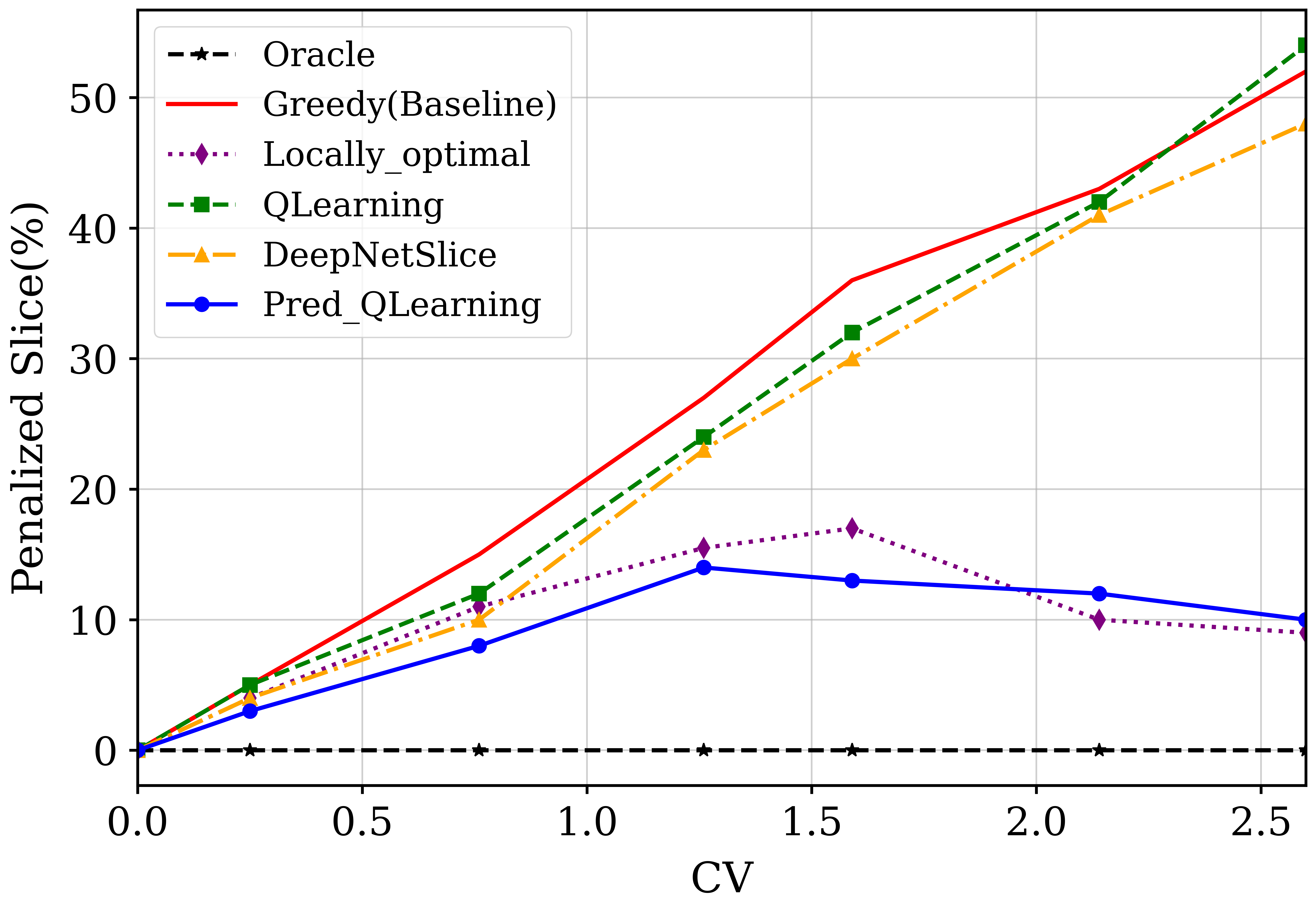}\vspace{-2em}%
        \label{subfig:penalty}
    }%
    \subfloat[Normalized Revenue]{%
        \includegraphics[width=0.50\columnwidth]{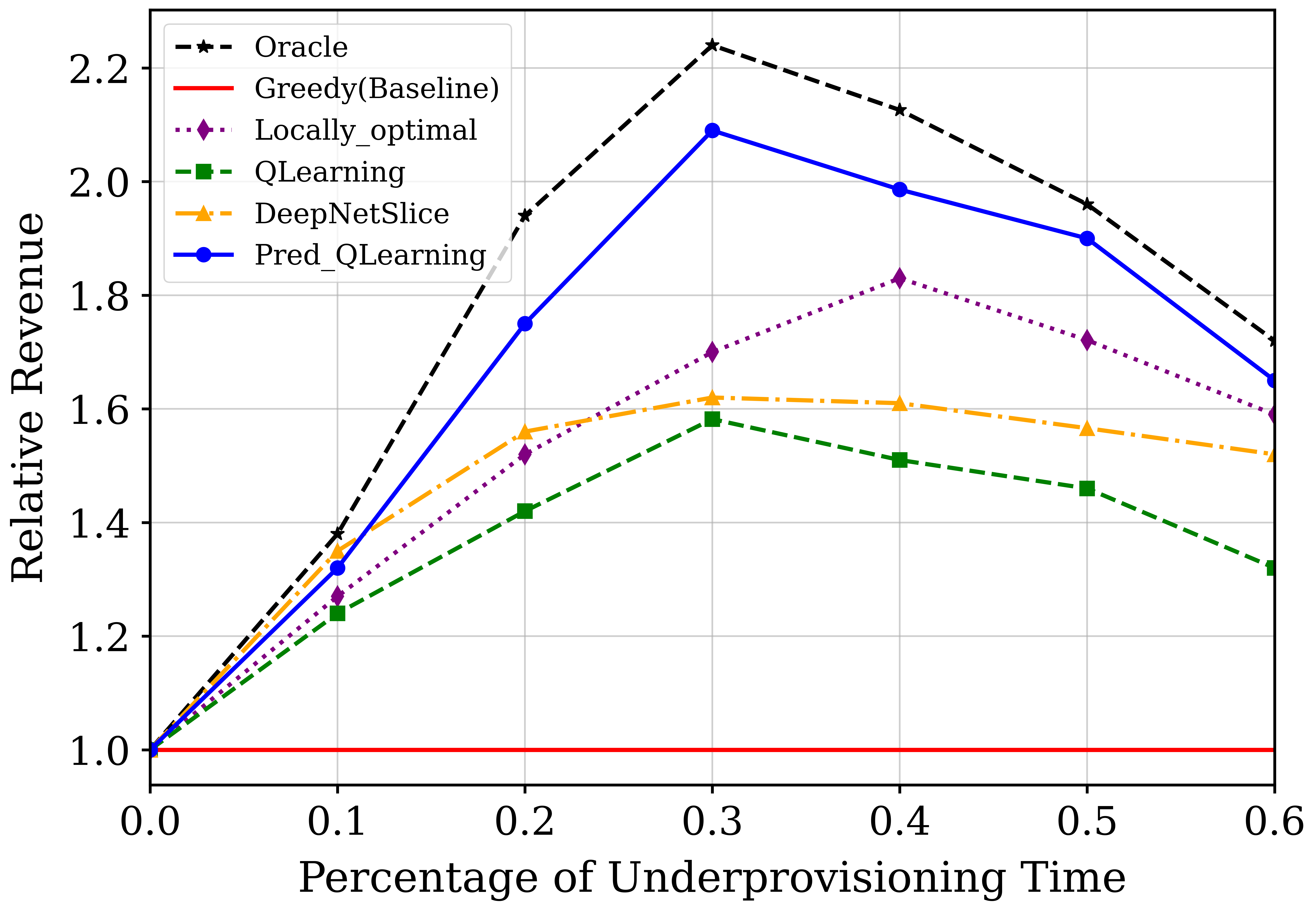}\vspace{-2em}%
        \label{subfig:underpro}
    }
    \caption{(a) Revenue of different SR algorithms as CV increases; (b) Same as (a) with revenue normalized with respect to Naive Greedy; (c) Percentage of SRs with negative revenue as CV increases; (d) Revenue as the maximum underprovisioning time fraction increases.}
    \label{fig:combined_revenue}
\vspace{-2em}
\end{figure}

\noindent\textbf{Impact of Link Fluctuation.} Fig.~\ref{subfig:average} shows the cumulative revenue for different SR algorithms under varying link CVs. 
In general, higher CV results in lower revenue due to added underprovisioning penalties. Fig.~\ref{subfig:random} normalizes revenue of the different SR algorithms with respect to the Naive Greedy baseline. For low CVs, all SR algorithms (except Random) achieve similar revenues. As CV increases, the revenue of Naive Greedy and Random become more similar. More importantly, for high CVs, predictive SR algorithms, namely Predictive Q-learning and Locally Optimal, can achieve revenues $3.5\times$ larger than Naive Greedy, while reactive SR algorithms, namely Naive Q-learning and DeepNetSlice, can achieve revenues $1.7\times$ larger than Naive Greedy. This significant difference between predictive and reactive SR algorithms is due to the tight integration of link capacity forecasting into SR admission control. Notably, Predictive Q-learning 
consistently outperforms all SR algorithms, with revenue that remains close to the (idealized) Oracle for all values of CV, suggesting that short-term capacity prediction combined with Q-learning can achieve  near-optimal performance. 

To further highlight the benefit of predictive SR algorithms, Fig.~\ref{subfig:penalty} shows the percentage of admitted SRs whose cumulative penalties $P_{it}$ exceed their lump sum reward $R_i$. As CV increases, all reactive SR algorithms experience increasing penalty rates. In contrast, predictive SR algorithms maintain substantially lower penalty rates, suggesting that explicit prediction can significantly reduce costly QoS-violation penalties. 

\noindent\textbf{Impact of Underprovisioning.}  
Recall that each simulated scenario is associated with a maximum underprovisioning time fraction. 
A low underprovisioning fraction indicates a simulated scenario with sufficient resources. A high underprovisioning fraction signals overload. As shown in Fig.~\ref{subfig:underpro}, when the fraction is low, all policies can admit (almost) all SRs and achieve similar rewards. However, when this fraction is between 0.2 and 0.4, (i.e., the operator has moderate resource availability), Predictive Q-learning greatly outperforms others by making better admission decisions. For high underprovisioning fractions, where resources are extremely scarce, all SR algorithms (including Oracle) degrade due to limited capacity. 


In~\cite[Appendix~\ref{app:prediction_implement}]{ourArXiv}, we discuss additional simulation results that evaluate predictive SR algorithms under different piecewise-linear penalty functions and when integrated to different RSL prediction algorithms. The results suggest that the proposed Predictive Q-learning algorithm consistently outperforms other SR algorithms across diverse scenarios. 



\section{Conclusion} \label{sec.conclusions} 
This paper addressed proactive slice admission control in mmWave x-haul, where weather-induced attenuation makes future capacity uncertain and can turn purely reactive admission into costly QoS-violation penalties. We introduced \system, a tightly coupled framework that (i) forecasts short-term link capacity as probabilistic distributions and (ii) exploits these distributions in an admission controller to balance immediate slice rewards against anticipated underprovisioning risk. Using real-world measurements from the NYC Mesh deployment and slice requests derived from residential traffic traces, our evaluation showed that incorporating capacity uncertainty into admission decisions yields substantial gains, particularly in volatile conditions. Looking forward, \system opens several directions for network optimization under uncertainty: extending from a star-like topology (as in Fig.~\ref{fig.design1}) to general multi-hop networks, and integrating joint demand and capacity forecasting.

\bibliographystyle{IEEEtran}
\bibliography{bibs/ref}

\appendices

\section{Notation and Clarification}\label{app:notation}
Table~\ref{tab:notation} summarizes the key notation used in this work. Additionally, we elaborate on several technical details and definitions from the paper.
\subsubsection{Training vs.\ Inference Clarification} \noindent\textbf{Training and Inference Notation.}
During training, samples $(\mathbf{X}, \mathbf{Y}) \in \mathcal{D}$ are indexed using a
relative prediction horizon $h = 1,\ldots,H$.
Accordingly, the model produces
$
\hat{\mathbf{Y}} = \{(\hat{\mu}_h, \hat{\sigma}_h^2)\}_{h=1}^{H},
$
where $h$ denotes the offset relative to the input window.

At inference time (Sec.~\ref{sec:predictor}), predictions are generated online.
At each evaluation time step $t$, the model outputs
$
\hat{\mathbf{Y}}_t = \{(\hat{\mu}_{t+h}, \hat{\sigma}_{t+h}^2)\}_{h=1}^{H},
$
which correspond to absolute future time indices.
These predictions are used immediately by \system to make admission control decisions.

\noindent\textbf{Teacher Forcing.}
In~\eqref{eq:attirnn_decoder}, AttIRNN uses teacher forcing: during training, $\hat{\mu}_{h-1} = y_{h-1}$ (ground truth), while at inference, $\hat{\mu}_{h-1}$ is the previous prediction. This allows the model to optimize all $H$ steps jointly during training while generating predictions sequentially at inference.

\noindent\textbf{Markov Approximation in Capacity Space.} For capacity propagation (Eq.~\eqref{eq:markov_recursion}), we use first-order Markov approximation for $h > 1$: $P(C_{t+h} | C_{t+h-1}, \hat{\mu}_{t+h}, \hat{\sigma}_{t+h}) \approx P(C_{t+h} | C_t, \{\hat{\mu}_{t+i}, \hat{\sigma}_{t+i}\}_{i=1}^{h})$. This approximation is both computationally tractable and well aligned with the ACM mechanism. In particular, ACM state transitions are governed by hysteresis threshold crossings relative to the current capacity state.
The predicted mean and uncertainty $(\hat{\mu}_{t+h}, \hat{\sigma}_{t+h})$
capture the likelihood of these threshold crossings and therefore provide sufficient information for estimating future capacity transitions.

\section{Dataset Description and Availability} \label{app:data_availability}
We use two publicly available datasets to support reproducibility and enable future work~\cite{jacoby2025openmeshdata,Yu2025Residential}.

\vspace{0.5ex}
\noindent\textbf{Wireless Measurements (OpenMesh):} To capture weather-induced attenuation across a city-scale mmWave wireless network, we utilize the OpenMesh dataset~\cite{jacoby2025openmeshdata}, comprising one-minute RSL logs from NYC Mesh's wireless network. NYC Mesh\footnote{NYC Mesh Community Network.} is a community-driven wireless network deployed in NYC. The dataset is hosted on Zenodo and formatted according to the OpenSense CML v1.1 NetCDF specifications.\footnote{M. Fencl \textit{et al.}, ``OpenSense: Data Formats and Conventions for Commercial Microwave Link Data v1.1,'' 2024.} Extended details regarding the network topology, data formats, and standardization are provided in~\cite{jacoby2025openmeshdata}.

\vspace{0.5ex}
\noindent\textbf{Residential Traffic for Slice Modeling:} To generate realistic slice requests, we use residential traffic traces derived from 520 NYC residences~\cite{Yu2025Residential}. These traces capture genuine user behavior, including video streaming and peer-to-peer activity. We map raw traffic flows to representative 5G service categories (eMBB, URLLC, and BE) and extract empirical distributions of throughput and duration.


\subsection{mmWave Network infrastructure}\label{app:network}
In this appendix, we describe the NYC Mesh network infrastructure used for our analysis and the methodology for link selection, expanding on the dataset overview provided in Section~\ref{sec:data}.

\vspace{0.5ex}
\noindent \textbf{Network Topology and Geography.}
Our dataset is derived from the NYC Mesh network, a community-owned urban network aiming to provide affordable internet service. NYC Mesh employs a hybrid tiered architecture blending mesh and star topologies, where fiber-connected Supernodes and wirelessly interconnected Hub Nodes form a redundant backbone to distribute connectivity to individual Member Nodes.
The infrastructure is concentrated in the dense urban environment of NYC (specifically Brooklyn and Lower Manhattan), as visualized in Fig.~\ref{fig.design1}. This deployment offers a unique testbed for analyzing weather effects typical of metropolitan areas, with further details provided in the OpenMesh dataset~\cite{jacoby2025openmeshdata}. 

\begin{figure}[htbp]
    \centering
    \includegraphics[width=0.9\linewidth]{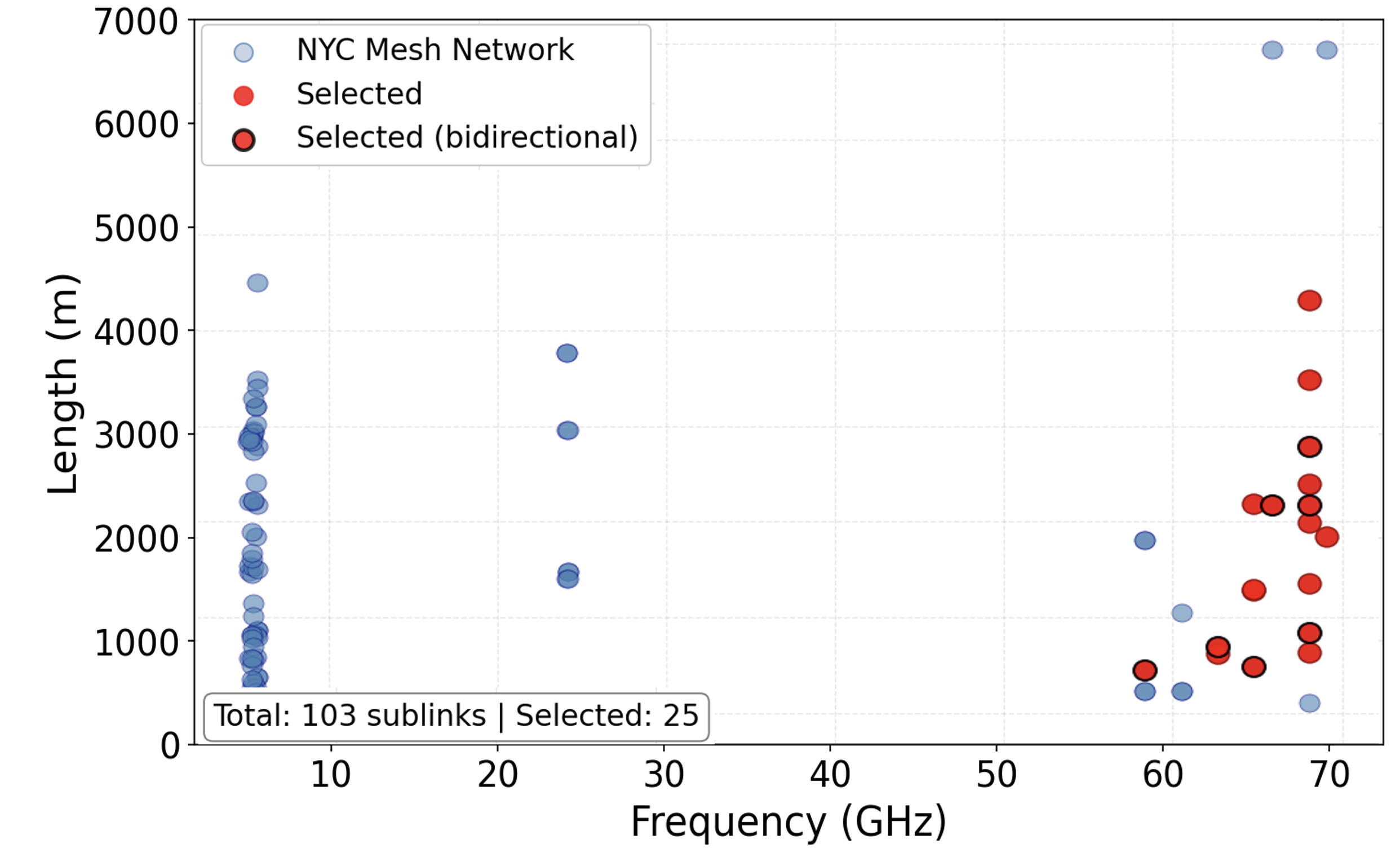}
    \caption{Scatter plot of link characteristics from the NYC Mesh network. The distribution shows path length versus operating frequency for all 103 available sublinks. The selected subset (red markers) isolates the used V-band links in this work.}
    \label{feature_scatter}
\end{figure}

\vspace{0.5ex}
\noindent\textbf{Data Subset and Link Selection.}
From the OpenMesh dataset~\cite{jacoby2025openmeshdata}, we constructed a targeted subset focusing strictly on V-band (60--70~GHz) mmWave links. This selection exploits the heightened sensitivity of these frequencies to atmospheric attenuation, which is advantageous for opportunistic weather sensing where lower frequencies often fail to capture light precipitation. Furthermore, analyzing these bands is critical for ensuring the high availability required by next-generation 5G/6G networks. The final subset was filtered for consistent reporting, with the resulting distribution of path lengths and frequencies illustrated in Fig.~\ref{feature_scatter}.

\subsection{Slice Request Generation from Network Traces} \label{app:slices}
As discussed in Sec.~\ref{sec:slice}, we used network traces collected from 520 Columbia University residential units to generate the SRs utilized in the evaluation of \system in Sec.~\ref{sec:evaluation}. We will detail the mechanism we developed to generate SRs from user demand network traces.

\vspace{0.5ex}
\noindent \textbf{Step 1. Collect and anonymize network packet traces.} 
Packets from/to the 520 off-campus residential units are mirrored from several aggregation switches managed by Columbia University to our servers in a nearby lab. We followed established practices to anonymize sensitive information such as IP addresses and MAC addresses (more details about the network topology and data collection can be found in \cite{Yu2025Residential}).

\vspace{0.5ex}
\noindent \textbf{Step 2. Identify flows for eMBB, URLLC, and BE slices.} 
Following established practices \cite{Yu2025Residential}, we aggregate packets into flows and identify services accessed by flows. 
A flow is defined as the packets with the same 5-tuple (i.e., source IP, destination IP, source port, destination port, transport protocol) within an hour. We associate flows with the query domain of the most recent DNS packet with corresponding IP addresses. We also associate TLS flows with their SNI fields. We also annotate flows with their corresponding service when possible, based on DNS/SNI domains, ports, destination IP addresses, and transport protocols. We begin with a curated list of approximately 200 keywords from DNS/SNI domains for 100 popular services. For example, when we see `nflx' in the associated domain, we know that the flow is used for Netflix service. We also use well-known port numbers and ASes to infer the associated services.

As there are no universally accepted industry standards or real-world implementations of network slicing, we identify widely used services that align with the characteristics of eMBB, URLLC, and BE and treat them as representative use cases for the three slicing categories. For the purpose of this study, we include only traffic flows caused by these selected services. For each flow, we know its start time, duration, average throughput, and the service it is used for. 

\vspace{0.5ex}
\noindent \textbf{Step 3. Determine slice types based on the throughput and duration statistics.} 
We assume that, for management simplicity, an infrastructure provider offers only a limited set of slice types, similar to how cloud providers today offer predefined virtual machine options. We further assume that providers will select reasonable options for slicing throughput and duration based on real traffic statistics. Therefore, for this study, we use three duration options--the 90th percentile of flow duration for each of the three slicing service categories. For throughput options, we use the 25th, 50th, 75th, and 95th percentiles of aggregated throughput for flows with the same start time and duration, as tenants are likely to group those flow requests from users when requesting slicing resources. These throughput values are scaled up by a factor of 4× to ensure that demands are aligned with the available link capacity in our simulation, and to anticipate increased bandwidth consumption due to the growing adoption of data-intensive services. The required throughputs for all SRs in our measurement are 0.4, 8.8, 19.2, and 27.2 Mbps. In total, we have 12 slice types.

\vspace{0.5ex}
\noindent \textbf{Step 4. Generate slices.}
Finally, we generate slices in a manner consistent with expected tenant behavior. A tenant receives flow requests from users, each specifying the service type and required network resources; however, tenants do not know in advance how long a user will maintain a flow, similar to how today’s services lack prior knowledge of session duration. Upon receiving a flow request, the tenant attempts to use existing slicing resources efficiently by placing the flow into an active slice with sufficient available capacity. If no such capacity exists, the tenant requests a new slice to accommodate the flow. Additionally, if a slice expires while the user is still using the service (i.e., the flow has not yet ended), we treat the remaining portion as a new flow request.

\begin{figure}[t]
\centering
\includegraphics[width=\columnwidth]{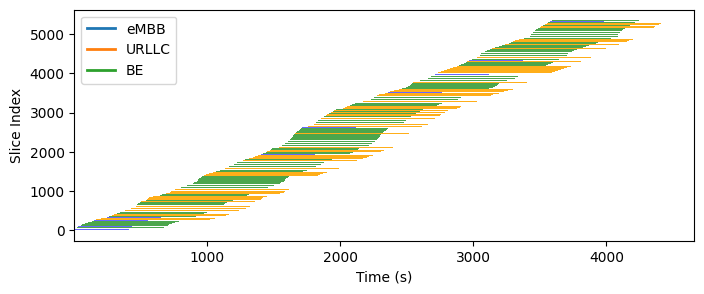}
\caption{An example of what slices look like over a one-hour interval. Each line denotes the start and end of a slice, and each color corresponds to a slicing service category.}
\label{fig.slicingExample}
\vspace{-1em}
\end{figure}

For one month of data, we use traffic collected from the same hour of each day (10am-11am) to ensure consistent traffic behavior. For each one-hour traffic trace, we generate slices and treat the result as a single slicing instance for simulation. In total, this yields 30 slicing instances used in our evaluation. We show an example of what slices look like in Fig.~\ref{fig.slicingExample}. X axis represents time, and Y axis represents the slice index. Each line denotes a slice, showing the start and end of a slice. Each color represents a slicing service category.

\section{Convex Piecewise-Linear Penalty Function} \label{app:penalty}
This appendix details the penalty model used to quantify service degradation
when admitted slices are underprovisioned. As discussed in Sec.~\ref{sec:formulation}, we represent the penalty for
underprovisioning SRs using a convex piecewise-linear function. Specifically,
for each active slice $i$ and time slot $t$, the penalty is defined as
$P_{it}=\max_{k\in\{1,\ldots,m\}}\{a_{s(i),k}(1-f_{it})+b_{s(i),k}\}$, where
$f_{it}\in[0,1]$ denotes the fraction of requested throughput allocated to the
slice. This construction preserves convexity and captures increasing service
degradation as the allocated fraction decreases. The coefficients satisfy
$a_{s,k}\ge 0$ and $b_{s,k}\le 0$ and are chosen such that $P_{it}=0$ when $f_{it}=1$.

\begin{figure}[t]
    \centering
    \includegraphics[width=0.85\columnwidth]{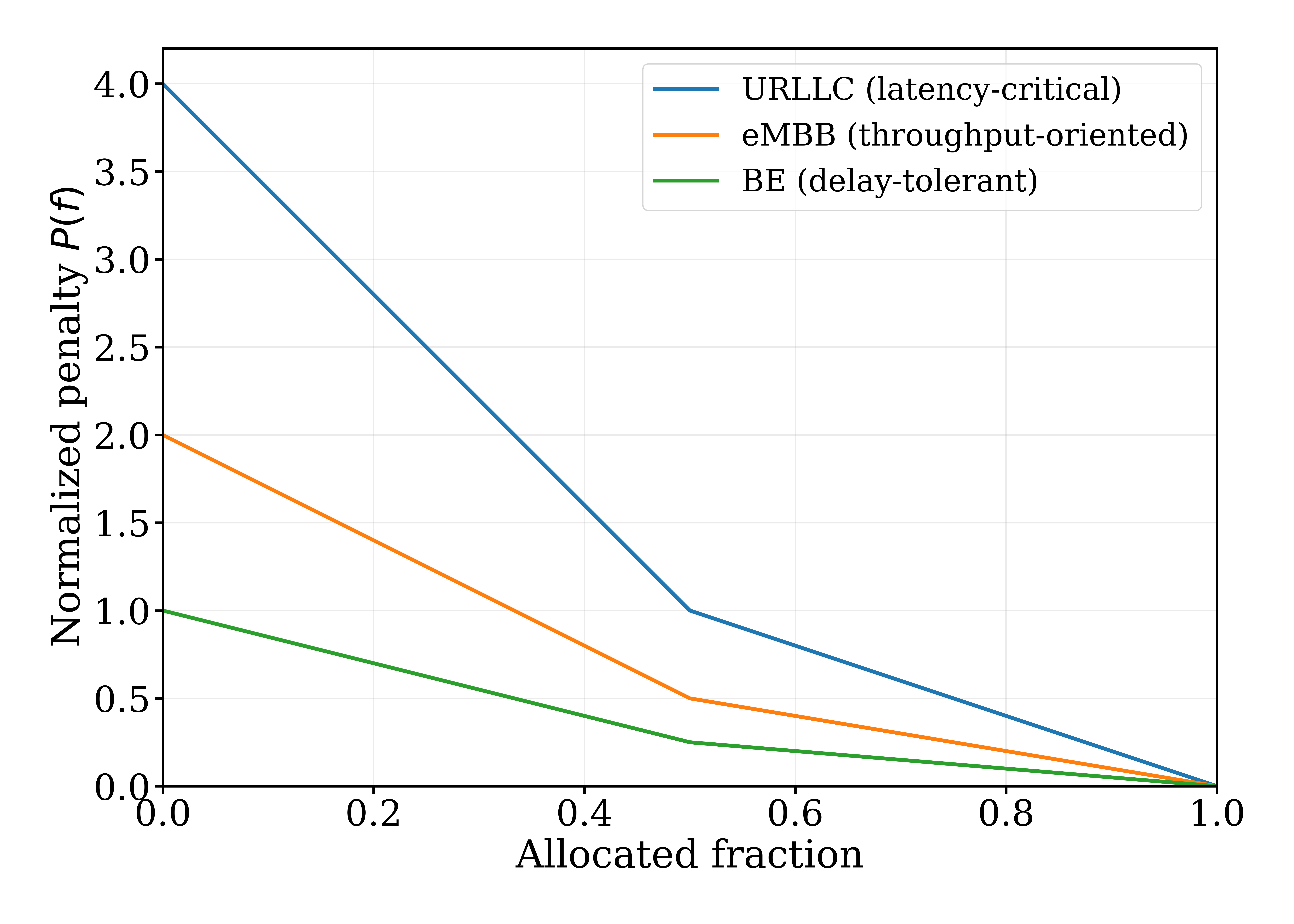}
    \caption{Penalty $P_{it}$ versus allocated fraction for different service types.}
    \label{fig:penalty_vs_fraction}
\vspace{-1em}
\end{figure}

In the evaluation of \system (Sec.~\ref{sec:evaluation}), we implement the penalty
model using service-type–dependent penalty functions via a standard epigraph reformulation:
\begin{equation}
\begin{aligned}
P_{it} &\ge a_{s(i),1}(1-f_{it})+b_{s(i),1} = \kappa \rho_i(1-f_{it}),\\
P_{it} &\ge a_{s(i),2}(1-f_{it})+b_{s(i),2}=\kappa \rho_i(2-3f_{it}),
\end{aligned}
\qquad \forall i,t.
\end{equation}
We use two linear segments ($m=2$) for all service types. $\rho_i$ denotes the price per throughput unit per slot of slice $i$, and $\kappa > 0$ is a global penalty severity coefficient that controls how strongly
the operator penalizes underprovisioning. Larger values of $\kappa$ correspond
to more conservative operation with higher sensitivity to QoS violations. The coefficients
$(a_{s,k}, b_{s,k})$ capture different tolerance levels to
underprovisioning across service types. Latency-critical services incur steeper penalties than best-effort traffic. Fig.~\ref{fig:penalty_vs_fraction} illustrates the resulting penalty functions used in Sec.~\ref{sec:evaluation}. All penalties satisfy $P_{it}=0$ when $f_{it}=1$ and increase monotonically as $f_{it}$ decreases, with a steeper slope under severe underprovisioning.

\section{Standard Hysteresis-based Adaptive Coding and Modulation (ACM)} \label{app:sig_cap}
As introduced in Sec.~\ref{sec:predictor}, we utilize a standard hysteresis-based ACM mechanism to transform RSL values into discrete link capacities. While the principles of ACM are well-established in wireless communications, this appendix details the specific RSL thresholds and Modulation and mapping parameters employed in our analysis.


\subsubsection{Predictive ACM Mapping} \label{app:ACM}
\begin{figure}[!t]
\centering
\includegraphics[width=\linewidth]
{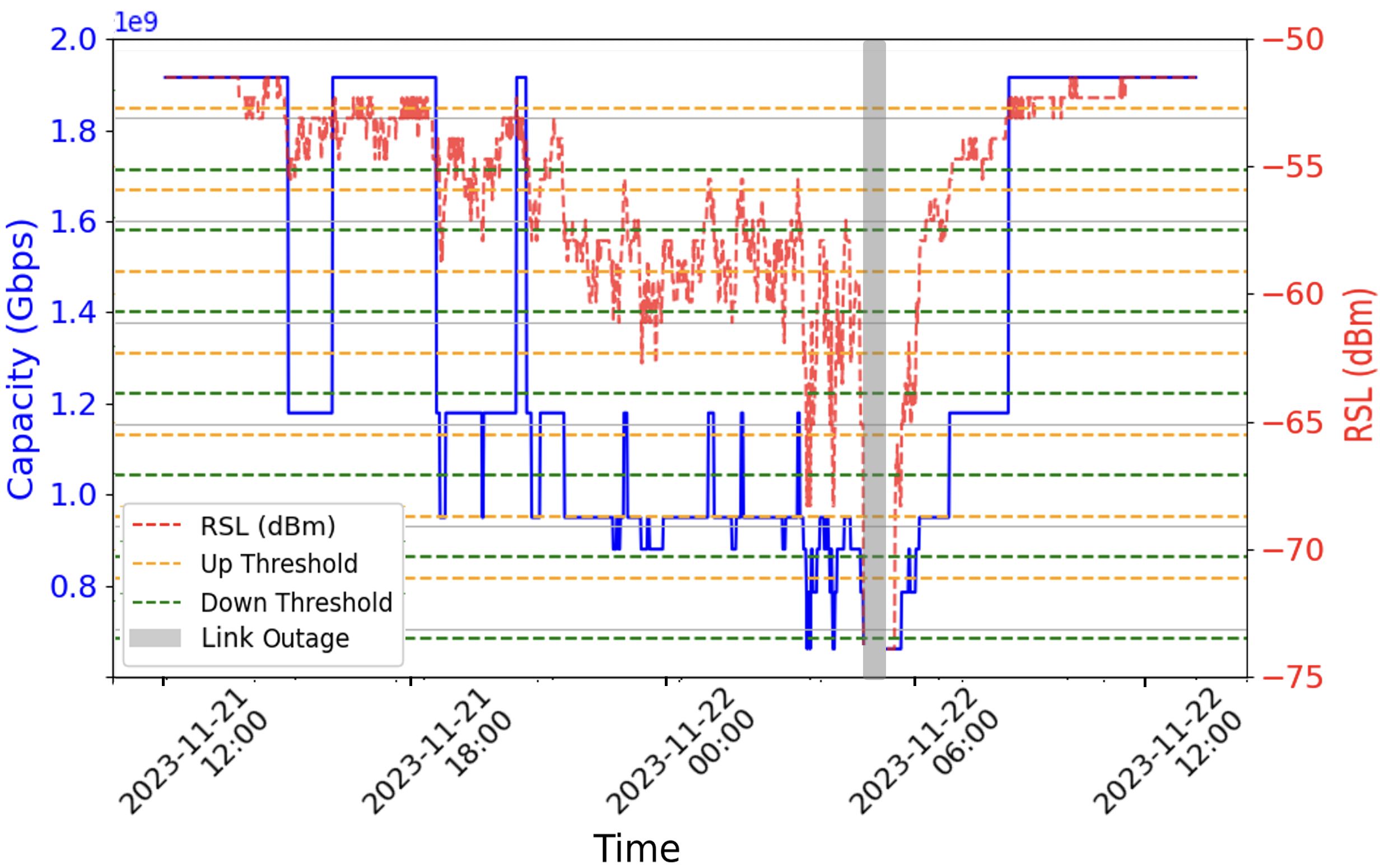}
\caption{Capacity levels during rain event (af60lr, Nov 21-22, 2023). RSL $\to$ capacity conversion via hysteresis (Table~\ref{table.cap}).}
\label{fig.cap_levels}

\end{figure}

Fig.~\ref{fig.cap_levels} illustrates the impact of a rain event on signal levels, highlighting the transitions between discrete capacity levels $\{C_l\}$ governed by hysteresis thresholds $(C_l^{\downarrow}, C_l^{\uparrow})$. For instance, an RSL drop below $-53.5$~dBm triggers a reduction to level $C_6$, a state that persists until the signal recovers above $-51.5$~dBm.
This mechanism follows standard ACM principles~\footnote{L. Bao, J. Hansryd, T. Danielson, G. Sandin, and U. Noser, “Field trial
on adaptive modulation of microwave communication link at 6.8 GHz,” in Proc. EuCAP, 2015, pp. 1–5}, adapting the link to the highest sustainable Modulation and Coding Scheme supported by the instantaneous RSL; specifically, we map these thresholds directly to discrete capacity values corresponding to standard modulation levels to focus on the resulting realizable throughput.
Crucially, we extend this mapping to the predictive domain—a concept for proactive routing explored in \cite{kadota2022switching}. However, unlike \cite{kadota2022switching}, which typically relies on point estimates, we apply the ACM thresholds directly to the \textit{entire} predicted RSL distribution. We model the future RSL as $y_{t+h} \sim \mathcal{N}(\mu_{t+h}, \sigma_{t+h})$, where the parameters are optimized via NLL minimization during training. By propagating these probabilistic forecasts through the discrete thresholds, we compute the explicit probability of each capacity level being available, providing a risk-aware input for our proactive admission control.

\noindent \textbf{Derivation of Thresholds and Capacity Values.} \label{sec.cap_derive}
Table~\ref{table.cap} lists the discrete capacity levels $\{C_l\}$ and hysteresis thresholds $(C_l^{\downarrow}, C_l^{\uparrow})$ utilized in our ACM scheme. To ensure a robust evaluation, we standardized these parameters by grouping the 68--69~GHz links into two representative clusters based on device type and path length:
(1) \textbf{AF60-LR} (1.8--2.2~km, 4 links), max capacity $C_7=1.95$~Gbps; and
(2) \textbf{Wave} (0.8--1.4~km, 2 links), max capacity $C_7=1.0$~Gbps.

Because raw capacity logs often lack discrete consistency, we derived the standardized mapping empirically by analyzing a single representative link from each cluster.
For each RSL interval, we assigned the capacity level $C_l$ corresponding to the \textit{empirical mode} of that representative link's historical performance.
To apply this standard mapping across the remaining links in each cluster, we normalized their signal scales to neutralize installation-specific variations (e.g., antenna alignment).
Specifically, we aligned the nominal (clear-sky) RSL for every link to a standardized reference point: 4~dB above the first degradation threshold $C_L^{\downarrow}$ (the point where the link drops from maximum capacity).
This adjustment ensures that the modeled ACM transitions are driven strictly by dynamic atmospheric attenuation rather than static link budget discrepancies.

\begin{table}[!t]
    \centering
    \caption{Hysteresis Thresholds for Predictive ACM}
    \begin{tabular}{lrrrrrr}
        \toprule
        \textbf{Level} & \multicolumn{2}{c}{\textbf{Capacity (Gbps)}} & \multicolumn{2}{c}{\textbf{Level Up (dBm)}} & \multicolumn{2}{c}{\textbf{Level Down (dBm)}} \\
        \cmidrule(r){2-3} \cmidrule(r){4-5} \cmidrule(r){6-7}
        ($l$) & \textbf{af60} & \textbf{wave} & \textbf{af60} & \textbf{wave} & \textbf{af60} & \textbf{wave} \\
        \midrule
        7 & 1.95 & 1.00 & -- & -- & -52.5 & -59.5 \\
        6 & 1.20 & 0.94 & -49.5 & -56.5 & -55.5 & -62.5 \\
        5 & 0.97 & 0.88 & -53.5 & -60.5 & -59.5 & -64.5 \\
        4 & 0.90 & 0.67 & -57.5 & -63.5 & -63.5 & -67.5 \\
        3 & 0.80 & 0.42 & -61.5 & -65.5 & -67.5 & -70.5 \\
        2 & 0.67 & 0.2 & -65.5 & -68.5 & -71.5 & -73.5 \\
        1 & 0.2 & 0.15  & -69.5 & -71.5 & -75.5 & -76.5 \\
        0 & 0.00 & 0.00 & -72.5 & -73.5 & $-\infty$ & $-\infty$ \\
        \bottomrule
    \end{tabular}
    \label{table.cap}
\end{table}


\section{Evaluation Details} \label{app:prediction_implement}


\begin{figure}[!t]
\centering
\includegraphics[width=0.9\linewidth]{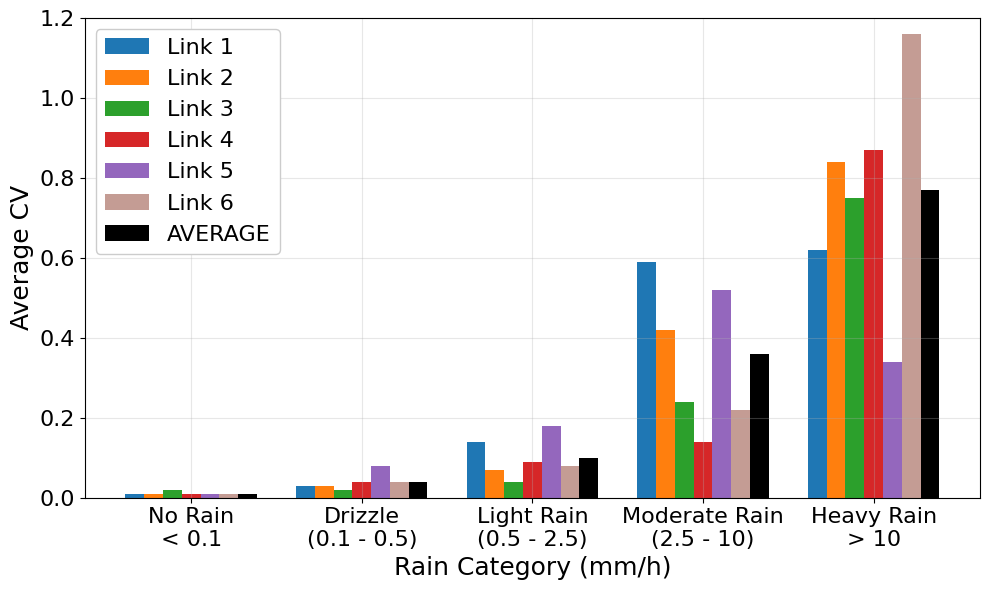}
\caption{Average CV categorized by rainfall intensity (WMO).}
\label{fig:rain_CV}
\vspace{-1em}
\end{figure}

\noindent\textbf{Prediction Modeling Strategy and Baselines.} 
To address multi-step short-term forecasting ($H=5$ min), we adopt a \textbf{univariate cross-learning framework}. While we initially explored multivariate architectures setup, including spatio-temporal attention \cite{kadota2022switching, jacoby2025spatio} and TSMixer \cite{chen2023tsmixer} with multi-link prediction objectives, our analysis indicated diminishing returns given our specific link topology and available measurements, ultimately favoring univariate frameworks. This aligns with findings that complex multivariate biases often induce overfitting when covariate correlations are limited, further discussed in \cite{chen2023tsmixer}. Consequently, for the neural evaluators (DeepAR, TSMixer, AttIRNN) in Sec.~\ref{sec:pred.evaluation}, we train a single global model utilizing a learnable \textbf{link-embedding} to capture link-specific dynamics without full multivariate overhead. 
For Neural Baselines, we adapt \textbf{TSMixer} \cite{chen2023tsmixer} and \textbf{DeepAR} \cite{salinas2020deepar} from their official PyTorch implementations. Specifically, we adopt the TSMixer-Ext architecture by concatenating flattened embeddings with time-series patches, while for DeepAR we utilize a NLL head for probabilistic regression to output distribution parameters. While in these neural models we freeze the trained weights for real-time deployment, the \textbf{Adaptive ARIMA} model, implemented via \texttt{pmdarima}, explicitly adapts to evolving patterns using a rolling-window approach to handle non-stationarity. Its optimal order is dynamically selected via AIC \cite{hyndman2008automatic}, with parameters refit every 10 minutes during evaluation. Despite these differing adaptation strategies, both frameworks successfully support real-time forecasting at the minute-scale resolution required for our evaluation.

\noindent\textbf{Data Split and Balancing Strategy:} To mitigate the inherent class imbalance where dry intervals vastly outnumber precipitation events, we prioritize rainy periods during training data selection. Training solely on raw chronological traces would bias the model toward predicting zero attenuation (the dominant class). We therefore employ standard statistical wet-dry classification methods to filter the dataset, ensuring a balanced representation of rain and non-rain conditions, adopting the multi-link methodology from \cite{jacoby2025spatio}. Notably, while this classification could be derived from external meteorological observations, we rely exclusively on signal-based statistics to preserve the system's stand-alone capability, avoiding dependence on auxiliary weather sensors.

\begin{table}[h]
    \centering
    \caption{\small Hyperparameter Grid Search Space and Configurations.}
    \label{tab:hyperparameters}
    \resizebox{\columnwidth}{!}{%
    \begin{tabular}{l|l|c|c|c}
    \toprule
    \textbf{Hyperparameter} & \textbf{Search Range} & \textbf{AttIRNN} & \textbf{DeepAR} & \textbf{TSMixer} \\ 
    \midrule
    \multicolumn{5}{l}{\textit{General Training Parameters}} \\ 
    \midrule
    Lookback Window ($T$) & $\{5, 10, 15, 20\}$ min & 15 min & 15 min & 20 min \\
    Batch Size & $\{64, 128, 256\}$ & 128 & 64 & 128 \\
    Learning Rate & $\{10^{-4}, 5 \times 10^{-4}, 10^{-3}\}$ & $10^{-3}$ & $10^{-3}$ & $5 \times 10^{-4}$ \\
    Dropout Rate & $\{0.0, 0.1, 0.2, 0.5\}$ & 0.1 & 0.1 & 0.5 \\
    \midrule
\multicolumn{5}{l}{\textit{Model Complexity Parameters}} \\
\midrule
    Hidden Dimensions & $\{32, 64, 128\}$ & 64 & 128 & 32 \\
    Layers / Blocks & $\{1, 2, 3, 4\}$ & 1 & 2 & 2 \\
    Link Embedding Dim. & $\{4, 8, 16\}$ & 8 & 8 & 4 \\
    \bottomrule
    \end{tabular}%
    }
    \footnotesize
    \textit{Note:} ``Layers/Blocks'' refers to encoder-decoder layers for RNN models and mixer blocks for TSMixer.
\end{table}

\noindent \textbf{Hyperparameter Optimization.}
All deep learning models were optimized via grid search to ensure fair comparison, selecting the configuration that minimized validation loss. To strictly prevent data leakage and preserve temporal ordering, the validation set was constructed using the final 10\% of the chronological data from the training links, rather than random shuffling. Table~\ref{tab:hyperparameters} outlines the hyperparameter spaces explored via grid search and the final configurations selected for each model. For architectural depth, we tuned the number of stacked components: encoder-decoder layers for AttIRNN variants, recurrent layers for DeepAR, and mixer blocks for TSMixer (denoted as "Layers/Blocks" in the tables).
To strictly isolate architectural contributions, ablation variants inherit AttIRNN's hyperparameters, whereas baselines are individually optimized for competitive fairness. 

\begin{table}[!t]
\centering
\scriptsize
\renewcommand{\arraystretch}{0.90}
\setlength{\tabcolsep}{4pt}
\caption{Prediction Performance across CV 
and horizon.}
\setlength{\tabcolsep}{2.5pt}
\label{tab:merged_table}
\begin{tabular}{llcccccc|cc}
\toprule
\multirow{2}{*}{\textbf{CV Range}} 
 & \multirow{2}{*}{\textbf{Step}} 
 & \multicolumn{2}{c}{\textbf{ARIMA}} 
 & \multicolumn{2}{c}{\textbf{DeepAR}} 
 & \multicolumn{4}{c}{\textbf{AttIRNN}} 
\\
\cmidrule(lr){3-4} \cmidrule(lr){5-6} \cmidrule(lr){7-10}
 & 
 & RMSE & $q_{\small.95}$
 & RMSE & $q_{\small.95}$
 & RMSE & $q_{\small.95}$
 & $\hat{\sigma}$ 
 & $\overline{\tilde{p}^{(1)}}$
\\
\midrule
\multirow{3}{*}{ \(  < 0.2 \)} 
 & t+1 & 0.30 & 0.87 & 0.38 & 0.98 & 0.32 & 0.82 & 0.18 & 0.96 \\
 & t+3 & 0.52 & 1.11 & 0.55 & 1.11 & 0.52 & 1.07 & 0.24 & 0.93 \\
 & t+5 & 1.34 & 1.44 & 0.64 & 1.36 & 0.58 & 1.25 & 0.39 & 0.91 \\
\midrule
\multirow{3}{*}{\(\ [0.2, 0.6] \)}
 & t+1 & 2.31 & 4.48 & 2.00 & 3.70 & 1.79 & 3.26 & 1.08 & 0.90 \\
 & t+3 & 3.81 & 6.40 & 3.16 & 5.76 & 2.88 & 5.18 & 1.28 & 0.84 \\
 & t+5 & 4.47 & 9.20 & 4.08 & 8.32 & 3.95 & 7.88 & 1.46 & 0.79 \\
\midrule
\multirow{3}{*}{\( >0.6\)}
 & t+1 & 2.83 & 6.33 & 2.58 & 5.94 & 1.93 & 5.56 & 2.00 & 0.87 \\
 & t+3 & 4.60 & 10.83 & 3.91 & 10.31 & 3.28 & 9.20 & 2.35 & 0.82 \\
 & t+5 & 5.22 & 13.71 & 4.45 & 12.48 & 4.16 & 10.84 & 2.60 & 0.72 \\
\bottomrule\end{tabular}
\vspace{-1.5em}
\end{table}

\noindent\textbf{Evaluation Setup.}
We evaluate our predictor and admission-control policy on real-world data from six V-band links excluded from the training set. The test dataset covers 34 hours across three distinct rain events in March–April 2024: \begin{itemize}[leftmargin=*,nosep] \item \textbf{Event I} (9 Mar, 03:15–13:45): Mean rain rate 5.45 mm/h, peak 21.6 mm/h, duration 10.5 h. \item \textbf{Event II} (23 Mar, 02:00–19:00): Mean rain rate 5.96 mm/h, peak 27.4 mm/h, duration 17 h. \item \textbf{Event III} (3 Apr, 10:00–16:00): Mean rain rate 5.80 mm/h, peak 45.7 mm/h, duration 6 h. \end{itemize}
The ground truth precipitation data was sourced from public archived datasets, averaging NOAA and Weather Underground records. We use CV-based metrics to normalize across diverse link characteristics (e.g., frequency, antenna type, path loss), rather than analyzing error sensitivity to individual features.

\noindent \textbf{CV-Based Evaluation.}
We use the CV to characterize link volatility independently of specific links or time periods. Higher CV indicates more frequent capacity fluctuations due to signal variations during intense rainfall. This enables fair comparison across heterogeneous links and varying weather conditions.
Fig.~\ref{fig:rain_CV} presents the average RSL CV for all evaluated links, grouped by rainfall intensity~\footnote{World Meteorological Organization, “Manual on Codes: International Codes, Volume I.1, Annex II to the WMO Technical Regulations,” World Meteorological Organization (WMO), Geneva, Switzerland, Tech. Rep. WMO-No. 306, 2017.}. 
Dry periods exhibit near-zero CV, indicating stable capacity, while CV increases sharply with rainfall intensity. Link~5 shows anomalously high CV under heavy rain due to signal outages.
In general, heavier rainfall yields higher CV values, while non-rain conditions show near-zero CV, indicating minimal capacity changes. 
Intense rain increases RSL fluctuations in the links, thereby reducing capacity a relationship shown in Fig.~\ref{fig:rain_event}.
It quantifies throughput stability, with low CV indicating consistent performance and high CV representing significant fluctuations which often require proactive admission control. 
Table~\ref{tab:merged_table} aggregates prediction results for ARIMA, DeepAR, and AttIRNN over the entire evaluation period, grouping them into three CV bands (low, medium, high) and three horizons (\(h = 1, 3, 5\)), with AttIRNN's confidence metrics also reported. It also reports AttIRNN's confidence metrics \(\hat{\sigma}_{t+h}\) and the mean $\overline{\tilde{p}^{(1)}}$ to quantify prediction uncertainty. When link variability is low (\(\text{CV} < 0.2\)), all models perform similarly; but at medium (\(0.2 \le \text{CV} < 0.6\)) and especially high variability (\(\text{CV} \ge 0.6\)), AttIRNN demonstrates a clear advantage, achieving the lowest RMSE and sharply reduced 95th-percentile absolute errors, indicating both higher accuracy and a lower frequency of significant prediction errors.

\begin{figure}[t]
\centering
\vspace{-1em}
\subfloat[Revenue vs. Penalty Coefficient]{
    \includegraphics[width=0.49\columnwidth]{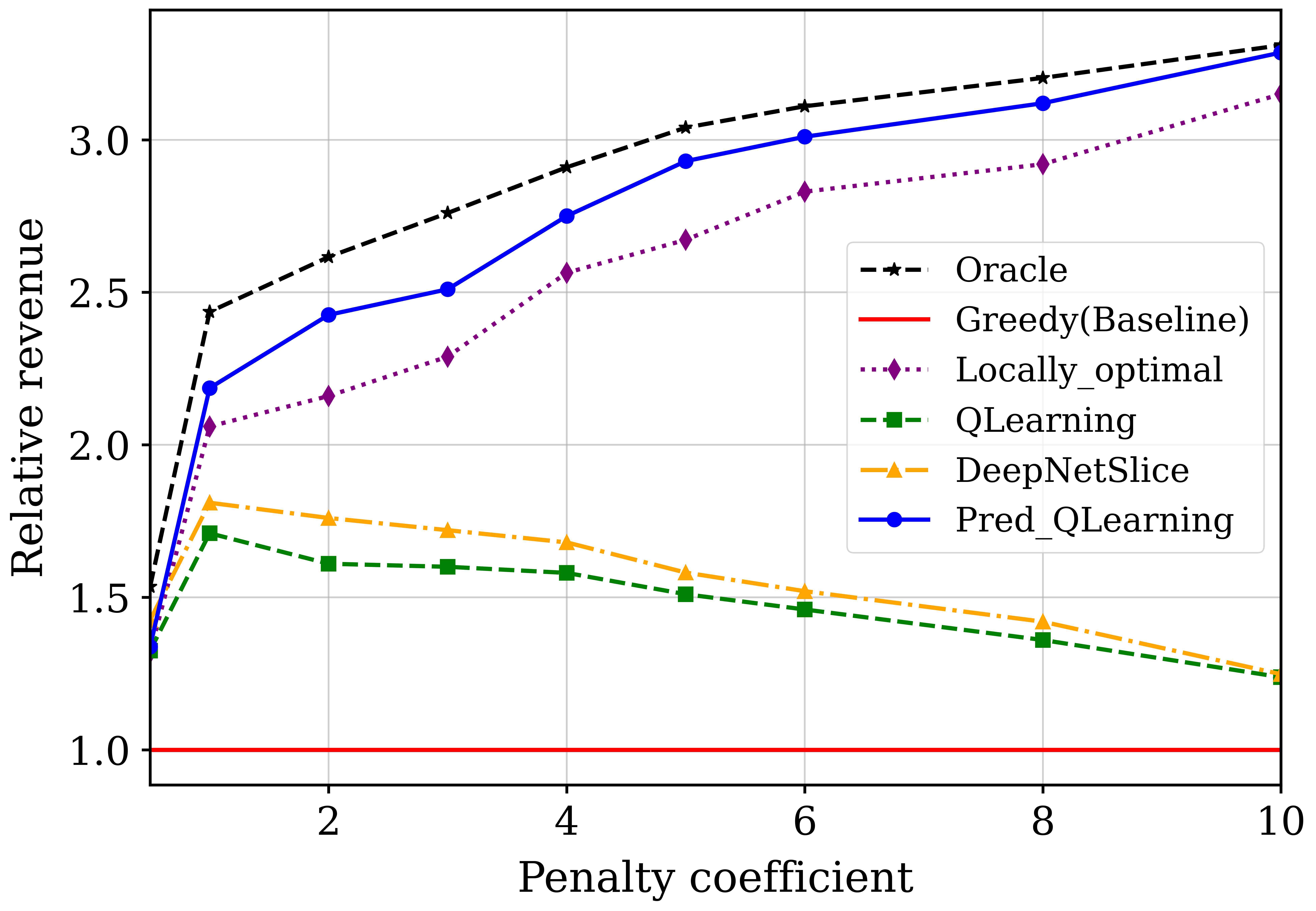}
    \label{subfig:coeffi}
}
\subfloat[Impact of Prediction Models]{
    \includegraphics[width=0.47\columnwidth]{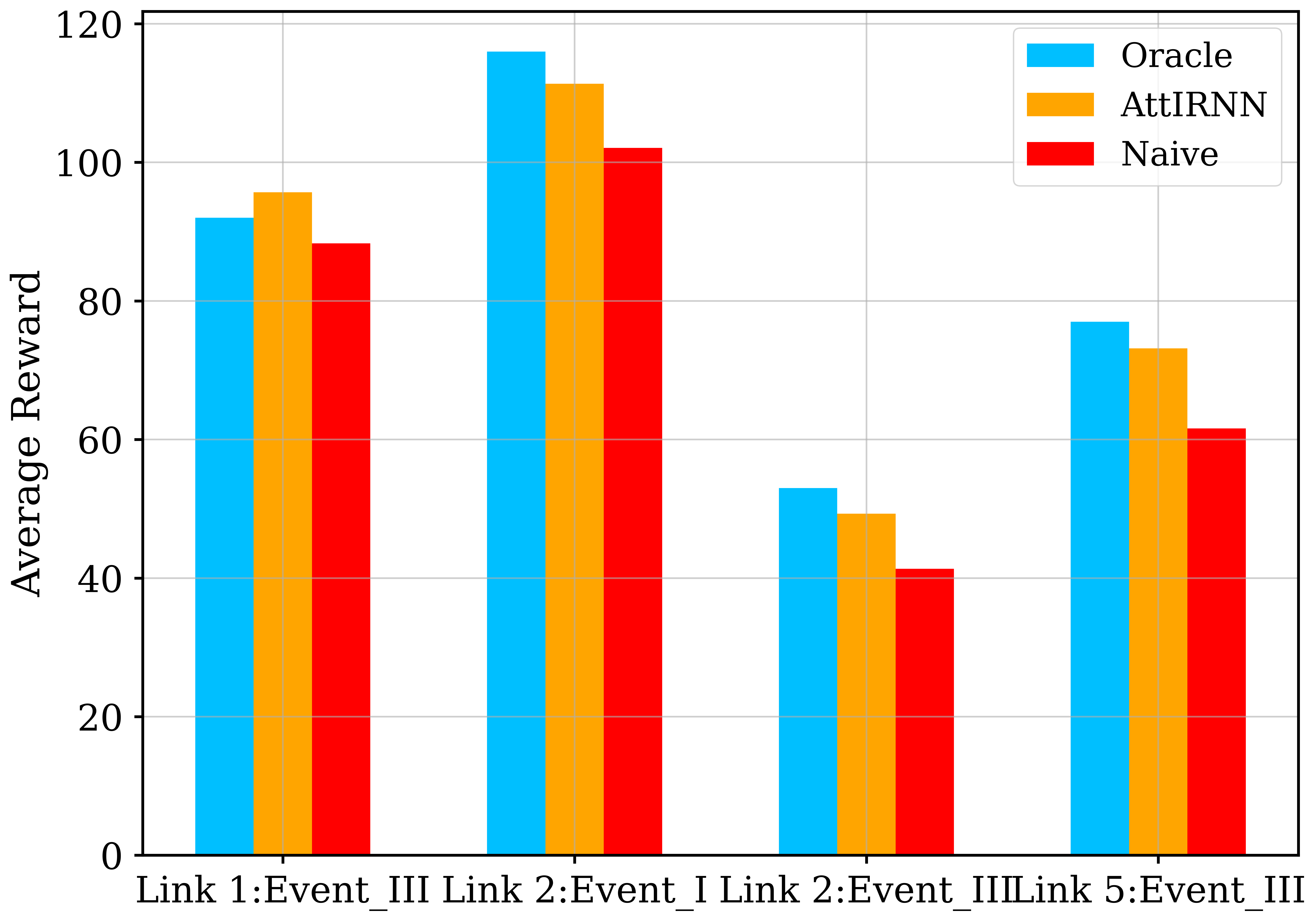}
    \label{subfig:models}
}
\caption{Revenue under different penalty coefficient or prediction models.}
\label{fig:revenue_2}
\vspace{-1em}
\end{figure}
\noindent\textbf{Impact of Penalty Severity Coefficient.} Fig.~\ref{subfig:coeffi} shows the impact of the penalty severity coefficient $\kappa$ on the relative revenue achieved by different algorithms as the penalty for underprovisioning increases. As the penalty coefficient grows, the performance gap among predictive models narrows, since all predictive approaches become increasingly conservative and tend to reject any slice that may introduce future penalties. Consequently, their admission decisions converge, leading to similar revenue outcomes. In contrast, the naive greedy, Q-learning, and DeepNetSlice models perform poorly as they fail to anticipate future capacity constraints and therefore incur substantial penalties.

\noindent\textbf{Impact of Prediction Models.} The previous results use AttIRNN as the capacity prediction model. The Naive model is easy to adopt in a system, assuming \(\hat{x}_{t+h} = x_t\), and serves as a baseline; despite its simplicity, it can perform reasonably well in noisy or volatile conditions. AttIRNN represents a more realistic learning-based predictor, as it has been shown to outperform other baseline models. The Oracle is used to evaluate the model with zero prediction error, serving as the ideal predictor benchmark. We evaluate the impact of these three prediction models on system performance. Fig.~\ref{subfig:models} shows that prediction accuracy influences the performance of predictive Q-learning. The Naive model tends to overestimate future capacity, which can lead to overly aggressive slice admissions and degraded performance due to increased underprovisioning penalties. AttIRNN achieves performance close to the Oracle on average. In particular, moderate underestimation of future capacity by AttIRNN can be beneficial, as it encourages more conservative admission decisions that reduce the risk of underprovisioning penalties under volatile network conditions.

\begin{table*}[h!]
\caption{Parameters and Decision Variables}
\centering
\begin{tabular}{|l|p{0.6\textwidth}|}
\hline
\textbf{Symbol} & \textbf{Description} \\
\hline
\multicolumn{2}{|c|}{\textbf{Perfect-Information MILP Parameters and Variables}} \\
\hline
$\mathcal{H}$ & Time-horizon. \\
$t$ & Time index $t\in\{1,\ldots,\mathcal{H}\}$. \\
$C_t$ & Capacity measurement at time slot $t$. \\
$N$ & Total number of SRs generated within the time-horizon. \\
$i$ & Chronological index of SR generation $i\in\{1,\ldots,N\}$. \\
$d_i$ & Throughput requirement of SR $i$. \\
$D_i$ & Duration of SR $i$. \\
$R_i$ & Lump sum reward for admitting SR $i$. \\
$P_{it}$ & Per-slot penalty incurred when SR $i$ is underprovisioned in slot $t$. \\
$t_i$ & Slot when SR $i$ is generated. \\
$z_i$ & Binary decision variable: 1 if SR $i$ is admitted at any time slot, 0 otherwise. \\
$\kappa$ & Global penalty severity coefficient. \\
$k_{it}$ & $1$ if slice $i$ is active at time slot $t$, $0$ otherwise. \\
$m_i$ & Number of intervals in the convex piecewise linear penalty function for SR $i \in \{1, \ldots, N\}$. \\
$a_{ik}, b_{ik}$ & Coefficient and intercept of the $k$-th linear segment, $k \in \{1, \ldots, m_i\}$, of the penalty function of SR $i \in \{1, \ldots, N\}$. \\
$f_{it}$ & Fraction of requested throughput assigned to SR $i \in \{1, \ldots, N\}$ at time slot $t \in \{1, \ldots, \mathcal{H} \}$. \\
$\rho_i$ & Price per throughput per time unit paid by a network SR. \\
\hline
\multicolumn{2}{|c|}{\textbf{Capacity Prediction (AttIRNN) Notation}} \\
\hline
$\mathbf{X}_{t}$ & The last-$T$ RSL samples $\{x_{t-T+1}, \dots, 
x_{t}\}$. \\
$T$ & Look-back window hyperparameter (number of historical RSL samples). \\
$\mathbf{Y}_{t}$ & The next-$H$ future RSL values $\{x_{t+1}, \dots, x_{t+H}\}$. \\
$\mathbf{x}_s$ & Static feature vector of link specifications. \\
$(\mathbf{X}, \mathbf{Y})$ & Generic $\mathbf{X}_{t}$, $\mathbf{Y}_{t}$ supervised input-output training pair from sliding window. \\
$\boldsymbol{\phi}_j^{E}$ & Encoder hidden states summarizing the input window. \\
$\mathbf{g}_s$ & Static embedding of link attributes $\mathbf{x}_s$ (learned vector representation). \\
$\boldsymbol{\phi}_h^{D}$ & Decoder hidden state at future step $h$. \\
$\mathbf{g}_h, \tilde{\mathbf{g}}_h$ & Attention-weighted encoder context and enriched context combining temporal-static features. \\
$\beta_{hj}, \tilde{\beta}_{hj}$ & Alignment scores and normalized attention weights, respectively. \\
$\hat{\mathbf{y}}_h$ & Stochastic prediction at horizon $h$: mean $\hat{\mu}_h$ and variance $\hat{\sigma}_h^2$. \\
$\hat{\mu}_{t+h}$ & Predicted mean RSL at time $t+h$. \\
$\hat{\sigma}^2_{t+h}$ & Predicted variance RSL at time $t+h$. \\
$C_l$ & Discrete capacity level ($l = 0, \ldots, L-1$). \\
$C_l^{\downarrow}, C_l^{\uparrow}$ & Lower and upper RSL thresholds for capacity level $C_l$. \\
$\Phi(\cdot)$ & The standard normal Cumulative Distribution Function (CDF). \\
$\tilde{p}_{h}(l' \rightarrow l)$ & One-step transition probability from capacity level $C_{l'}$ to $C_l$ at time horizon $h$. \\
$\tilde{p}_{l,h}$ & Estimated likelihood of the link operating at capacity level $C_l$ at time horizon $h$. \\
$H$ & Short-term time horizon for capacity prediction. \\
\hline
\multicolumn{2}{|c|}{\textbf{Slice Admission Control Notation}} \\
\hline
$s_{i(j)}$ & System state associated with SR $i(j)$, represented as
$(\mathbf{n}, \mathbf{A}_{\text{new}}, c_f)$. \\

$a_{i(j)}$ & Action taken for SR $i(j)$, defined as $a_{i(j)} \in \{0,1\}$ (reject or admit). \\
$\mathcal{N}_t$ & Set of active SR indices at time slot $t$. \\

$\mathcal{A}_t$ & Set of SRs arriving at time slot $t$. \\

$\mathbf{n}$ & Vector $\mathbf{n}=(n_1,\ldots,n_{12})$ recording the number of
active SRs of each type. \\

$\mathbf{A}_{\text{new}}$ & One-hot encoding vector indicating the type of the
currently considered SR $i(j)$. \\

$c_f$ & Predictive indicator of future capacity availability. \\

$Q(s,a)$ & Q-value function estimating the expected long-term revenue of taking
action $a$ in state $s$. \\

$r(s_{i(j)}, a_{i(j)})$ & Immediate reward obtained by taking action $a_{i(j)}$
in state $s_{i(j)}$. \\

$\lambda$ & Risk-penalty scaling coefficient. \\

$\alpha_j$ & Learning rate at iteration $j$, computed as
$\alpha_j = 0.5 / o(s_{i(j)}, a_{i(j)})$. \\

$o(s_{i(j)}, a_{i(j)})$ & Occurrence count of the state-action pair
$(s_{i(j)}, a_{i(j)})$. \\

$\epsilon$ & Exploration probability in the $\epsilon$-greedy policy. \\
\hline
\end{tabular}\label{tab:notation}
\end{table*}

\end{document}